\documentclass[a4paper,10pt]{article}
\usepackage{titling}
\usepackage{chngcntr}
\usepackage{graphicx}
\usepackage{hyperref}
\usepackage{booktabs}
\usepackage{multirow}
\usepackage[
top    = 1in,
bottom = 1in,
left   =1in,
right  = 1in]{geometry}
\usepackage{adjustbox}
\usepackage{lscape}
\usepackage{rotating}
\usepackage{epstopdf}
\usepackage{multirow}
\usepackage{longtable}
\usepackage{array}
\usepackage{footnote}
\usepackage{array}
\usepackage{multicol}
\usepackage{threeparttable}
\usepackage{footnote}
\usepackage{rotating}
\usepackage{tikz}

\usepackage{color}

\usepackage{booktabs, makecell, tabularx}

\newcolumntype{L}{>{\RaggedRight}X}

\usepackage[skip=0.5\baselineskip]{caption}
\counterwithin*{section}{part}

\usepackage[english]{babel}
\usepackage[protrusion=true,expansion=true]{microtype}	
\usepackage{amsmath,amsfonts,amsthm,amssymb}

\DeclareMathOperator{\EX}{\mathbb{E}}

\title{A Bloom Filter Survey: \\Variants for Different Domain Applications}		

\author{
  Anes Abdennebi\\
  \texttt{anesabdennebi@sabanciuniv.edu}
   \and
  Kamer Kaya\\
  \texttt{kaya@sabanciuniv.edu}
}
\date{}
\begin{document}
\maketitle

\begin{abstract}
There is a plethora of data structures, algorithms, and frameworks dealing with major data-stream problems like estimating the frequency of items, answering set membership, association and multiplicity queries, and several other statistics that can be extracted from voluminous data streams. In this survey, we are focusing on exploring randomized data structures called Bloom Filters. This data structure answers whether an item exists or not in a data stream with a false positive probability fpp. In this survey, many variants of the Bloom filter will be covered by showing the strengths of each structure and its drawbacks i.e. some Bloom filters deal with insertion and deletions and others don’t, some variants use the memory efficiently but increase the fpp where others pay the trade-off in the reversed way. Furthermore, in each Bloom filter structure, the false positive probability will be highlighted alongside the most important technical details showing the improvement it is presenting, while the main aim of this work is to provide an overall comparison between the variants of the Bloom filter structure according to the application domain that it fits in.
\end{abstract}

\title{Set Membership Sketches}
\author{}
\date{}

\section{Introduction} 
  Streams have many applications in today's data processing systems such as Internet of Things \cite{SINGH2018440}, Edge Computing \cite{SINGH2018440}, and Databases  \cite{Bloom:1970:STH:362686.362692} \cite{appDB-BF} \cite{appDB-BF1} \cite{CALDERONI20154} \cite{appDB-CS} \cite{Chen2014RobustSR} \cite{appDB-CS1}. A Bloom Filter~(BF) is a compact data structure which can answer membership queries on infinite streams having a large number of unique items~\cite{Bloom:1970:STH:362686.362692}. Compared to the traditional approaches, a Bloom Filter occupies much less space.
However, it is a probabilistic data structure in the sense that it allows false-positive responses. When an item is queried and BF responds with {\em true}, the item {\em may or may not} exist in the stream. On the contrary, when the response is {\em false}, the item does not exist in the stream.

  The applications of Bloom filters are many; they are frequently used in networks \cite{Bloom:1970:STH:362686.362692} \cite{Geravand:2013:SBF:2560974.2561537} \cite{doi:10.1080/15427951.2004.10129096} \cite{bloomierInNetworks} \cite{1210033} \cite{Ledlie:2002:SPS:1133373.1133397} \cite{Hsiao:2001:GRS:509506.509515} \cite{Rhea} \cite{1019229} \cite{916648} and web applications\cite{web-caching} \cite{web-app-bf}. 
  BFs are also used to track items stored in databases to reduce or avoid expensive and unnecessary accesses~\cite{Cormode:2017:DS:3134526.3080008}, for instance, an email service provider tries to minimize the emails database accesses to the lowest possible rate such that when a user tries to log in to his account, a query is performed in the used Bloom Filter to check whether the entered email exists or not, if it does, then another query is executed to check the validity of the entered credentials, otherwise, a negative and fast (since it queried the BF and not the database itself) response is returned to the user about the entered email existence. BF is useful too in spell-checking where the existence of a word in the dictionary is queried in the Bloom Filter to guarantee a fast response \cite{Bloom:1970:STH:362686.362692}. In biometric templates protection schemes, transferring the biometric data of the iris which is stored in matrices to Bloom Filters forms a secure scheme to prevent impersonation attacks, grants privacy protection for the subjects and individuals \cite{BioBF}, and also guarantees two intrinsic requirements, the \textit{irreversibility} and \textit{unlinkability} \cite{ISO/IEC24745:2011} \cite{GOMEZBARRERO201837} \cite{RATHGEB20141}. In cloud computing, the work proposed by \cite{cloud-comp} uses the \textit{Counting Bloom Filter} to perform searches over encrypted data in the cloud server and allows the data users to obtain proofs that the cloud server is indeed executing their queries and nothing wrong with its condition i.e. not compromised. Bloom Filters are also used in privacy preservation applications \cite{vatsalan2016multiparty}, for instance, in finding the records of an entity across databases hosted by different parties without putting it under the risk of compromising or exposure. This process is called \textit{Privacy-Preserving Record Linkage} and uses BF to ensure the privacy of the information \cite{schnell2009privacy}. In the same context of privacy preservation, another work \cite{calderoni2015location} proposes a new Bloom Filter variant called the \textit{Spatial Bloom Filter} to maintain users' privacy in location-aware applications. In machine learning, \cite{NIPS2013_5083} suggests a new variant of BF in \textit{Multilabel Classification } that handles large number of labels.



Although the original data structure is simple, there are many variants in the literature. In this work, we survey all the Bloom Filters and organize them into sections based on their main focus. These improvements or optimizations may target different aspects in the structure of BF like reducing the memory space used in the filter, improving the false positive rate, dealing with dynamic datasets, and reducing the computation overhead. Furthermore, there also exist studies to extend BFs for different applications and operations such as querying multiplicity, handing deletions, etc. We also cover these studies in this survey and provide an application-focused classification. In addition, we propose an open-source, high-performance, parallelized BF library which contains the implementations of all the existing proposals. We provide an extensive experimental evaluation of the existing studies in terms of their accuracy and performance on a single-node server as well as on single-board computers that usually need to use sketches for data processing due to their memory restrictions.

\section{Background and Notation}
A Bloom Filter uses a bit vector ${\tt bf}[.]$ of size $m$ and $k$ hash functions. Each hash function takes an element $e$ from the universal set $\cal U$ as an input and outputs a location in the bit vector. That is $h_i: {\cal U} \rightarrow \{0, \ldots, m-1\}$ for all $1 \leq i \leq k$. A BF has two main operations; \textit{insertion} \& \textit{querying}. Initially, all bits are set to $0$. To {\em insert} an item $x$, the bit positions $h_{1}(x), h_{2}(x),\ldots, h_{k}(x)$ are computed and the corresponding bits, ${\tt bf}[h_{1}(x)], {\tt bf}[h_{2}(x)], \ldots, {\tt bf}[h_{k}(x)]$ are set to $1$. To {\em query} an item's existence, the hash functions $h_{1}(x), h_{2}(x),\ldots, h_{k}(x)$ are computed again. If all the corresponding bits for these positions are $1$ BF returns {\em true}. Otherwise, it returns {\em false}. 
 

  \begin{figure}[h]
  \begin{center}
  \includegraphics[width=0.65\textwidth]{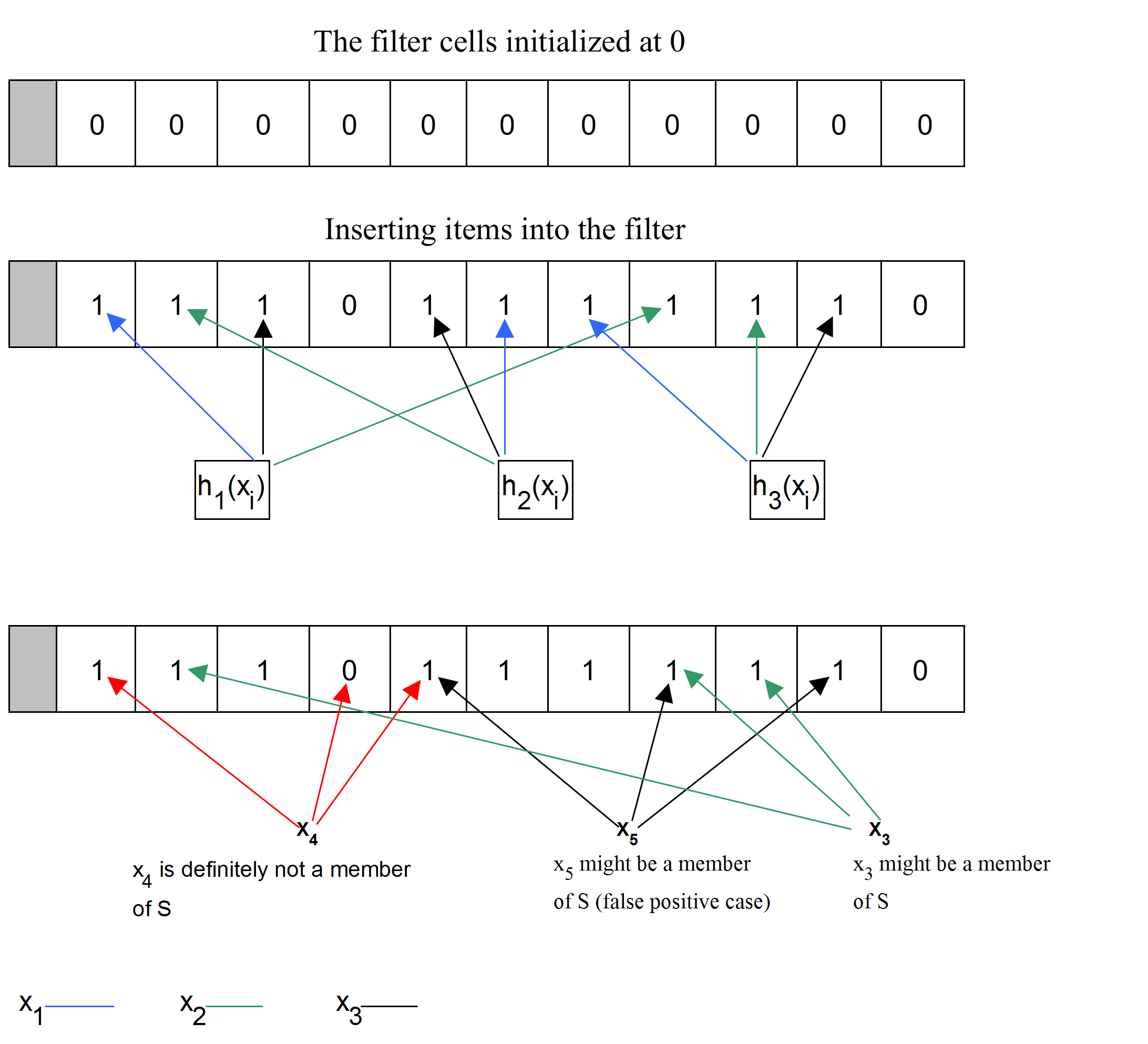}
  \end{center}
  \caption{A toy BF with $m = 11$ and $k = 3$. Three items are inserted to the BF; the hash function outputs for $x_{1}$ are  $0, 5$ and $6$. For $x_{2}$, they are $2, 4$ and $9$ and for $x_{3}$, they are $1, 7$ and $8$. Two extra items are queried; $x_{4}$ with hash positions $0$, $3$ and $4$ and $x_{5}$ with hash positions $4$, $7$ and $9$. 
  }
  \label{fig:Bloomfilter}
  \end{figure}
 
A toy BF with $m = 11$ bit vectors and $k = 3$ hash functions is shown in Figure~\ref{fig:Bloomfilter}. To query an item, the corresponding bit positions are checked; for $x_{4}$, these bits are ${\tt bf}[0], {\tt bf}[3]$ and ${\tt bf}[4]$. Since ${\tt bf}[3]$ is $0$, $x_{4} \notin S$. For $x_{5}$, although all the corresponding bits are $1$, this is a false positive response~(due to a hash collusion at ${\tt bf}[8]$). Querying the item $x_{3}$ would return "true" after checking that the bits in positions $1, 7$ and $8$ are all set to $1$ and therefore, $x_{3}$ is not in the filter.

 One of the main metrics to evaluate the effectiveness of a BF is the {\em false positive probability} which can give information about how accurate the BF is. Assuming the hash functions uniformly distributes the items to their range, the probability of having a given bit equal to 0 is  

\begin{equation} \label{1_fppv1}
p = \left(1-\frac{1}{m}\right)^{kn} \approx e^{-kn/m}
\end{equation}

\noindent where \textbf{$m$} is the vector size, $n$ is the number of unique items in the stream, and \textbf{$k$} is the number of hash functions used. Starting with $p$, the false positive probability is 

\begin{equation} \label{fpp_SBF}
\epsilon = (1-p)^{k} \approx \left(1-e^{-kn/m}\right)^{k}.              
\end{equation}

\section{Bloom filters focusing on reducing the fpp}

Several variants of Bloom Filter appeared lately aiming to reduce the false-positive rate in a Bloom Filter instance, some works propose solutions based on increasing the number of hash functions but in the cost of allowing false negatives with a certain probability, other works suggest increasing the bit vector size to avoid bits collisions while inserting the data stream elements, however, this method causes additional memory overhead.

\subsection{The yes-no Bloom Filter}
The yes-no Bloom filter's structure \cite{carrea2016yesno} resides on the same structure of the \textit{Standard Bloom Filter (SBF)} in answering the membership queries, moreover, it preserves information about the elements that triggers false positives in order to prevent them from occurring in future queries. The main structure of yes-no BF is composed of:
\begin{itemize}
	\item{yes-filter}: which works the same way as \textit{SBF}, it has a bit vector.
	\item{no-filter}: which encodes the elements that arouse false positives.
\end{itemize}
\textit{yes-no BF} uses a bit vector of size $m$, two sets of independent hash functions $H_{1 \leq i \leq k}$ and $G_{1 \leq j \leq k^{'}}$ such that $k$ and $k^{'}$ represent the number of hash functions used in each set respectively and $k^{'} < k$. The bit vector is split into two partitions; $p$ bits for the \textit{yes-filter} and $rq$ bits for $r$ \textit{no-filters} where $p >> q$.\\
For the insertion of an element $e$ in \textit{yes-no filter}, the $h_{i}(e)$ bits are set to $1$ in the \textit{yes-filter}, and for the construction of the \textit{no-filter}, the $g_{j}(e)$ bits are set to $1$, however, an element can be inserted only in one \textit{no-filter} from the $r$ ones.\\ During the querying process, an item $e$ is said to be \textit{"probably belongs to $S$"} if all the positions $h_{i}(e)$ are $1$ in the \textit{yes-filter} and there is no indication from the \textit{no-filters} that this item is a false positive. However, this method can lead to false negatives results.\\
The false positive probability for the \textit{yes-no BF} is composed of two parts; the \textit{$fpp_{yes-filter}$} and the \textit{$fpp_{no-filters}$} \\
First, the false positive probability of \textit{$fpp_{yes-filter}$}:
\begin{equation}
	fpp_{yes-filter} \approx \left(1-e^{-\frac{kn}{p}} \right)^{k}
\end{equation} 
Second, the false positive probability of \textit{$fpp_{no-filters}$}:
\begin{equation}
	fpp_{no-filters} \approx \left(1-e^{-\frac{k'n}{q}} \right)^{k'}
\end{equation} 
Then the false positive rate for the whole \textit{yes-no Bloom filter} is given by:
\begin{equation}
	fpp_{yes-no BF} = \left(1-e^{-\frac{kn}{p}} \right)^{k} \left( \left(1-e^{-\frac{k'n}{q}} \right)^{k'} \right)
\end{equation}

\begin{figure}[h]
  \begin{center}
  \includegraphics[width=0.65\textwidth]{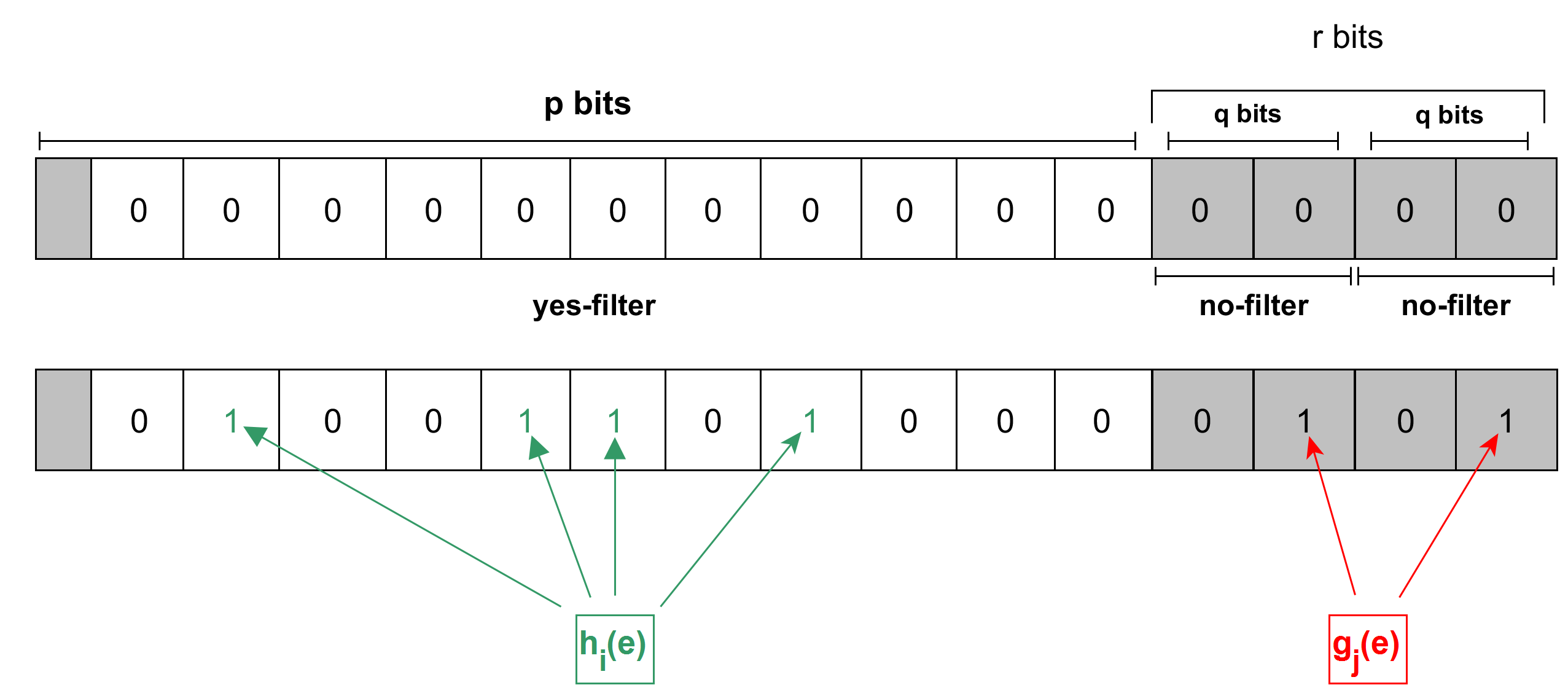}
  \end{center}
  \caption{The insertion of an element $e$ in the yes-no BF}
  \label{fig:yes-no BF}
  \end{figure}
  
\newpage

\noindent In Fig. \ref{fig:yes-no BF}, inserting the element $e$ is similar to the insertion process in BF, it inserts the element into the \textit{yes-filter} where the hash functions set $h_{i}$ is used to set the bit positions $1, 4, 5$ and $7$ to $1$. Whenever the \textit{yes-filter} is queried, the \textit{no-filters} of each element are modified (the corresponding bits are set to $1$) by storing the elements that cause false positives as it is shown in the figure.

\subsection{The Variable-Increment Counting Bloom Filter}
It is an enhanced version of the \textit{Counting Bloom Filter-CBF} concerning the accuracy of queries' answers (lower \textit{fpp}) and memory efficiency aspects. The Variable-Increment Counting Bloom Filter, known as \textit{VI-CBF} defines a list of possible variable increments called \textit{L} such that an incoming element would be hashed to one of those variables, then this variable's value can be added to the corresponding counter of the element, whereas, for deletions, the opposite operation is performed by decrementing the counters with the relevant variables from the list \textit{L}. And while querying and element $x$, each counter's hashed value in $L$ of the queried item is checked whether it is a factor in the sum or not, if it is not, then definitely $x$ is not a member of the set, otherwise, it might be (still can give a false positive). As an example to describe this data structure much better and to show the difference, in the matter of accuracy, between the regular \textit{CBF} and \textit{VI-CBF}, let's suppose that $L={2,3,4,5}$ which represents the variable increments, and there is an element $x$ to be added to the filter, $h_{i}$ where $(0 \leq i < k)$ represents the hash functions. So, when $x$ is being hashed to one of the values in $L$, let's say $5$, it will be added to the corresponding counter. When querying $z$, the membership query that will check whether the item exists or not will go over all the counters' hashed values and test whether it is a part of the sum in the current counter or not.

\begin{figure}[!h]
  \begin{center}
  \includegraphics[width=0.7\textwidth]{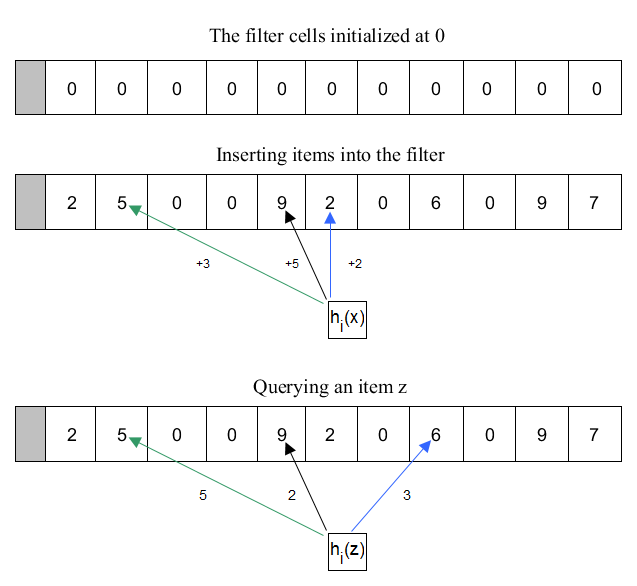}
  \end{center}
  \caption{Example demonstrating the work of VI-CBF}
  \label{fig:variable-increment BF}
 \end{figure}

\noindent As figure \ref{fig:variable-increment BF} shows, the second hashed value gave $2$ and since $9-2=7$ which is not included in $L$, therefore, the item $z \notin S$ because it does not seem to form any sum from the set $L$ i.e. $2+7=9 \notin L$.
  
\noindent The work \cite{6195563} Presents two schemes for based on \textit{CBF} to reduce the \textit{false positive rate} and the increase the memory efficiency; First, $B_{h}Scheme$, this scheme is based on the $B_{h} sequences$  \cite{Graham1996} and in a nutshell, this structure uses two counters, $c_{1}$ which increments by one, and $c_{2}$ which increments with the hashed value from $L$ every time an element is inserted to the filter. Moreover, $B_{h}-CBF$ uses two sets of $k$ hash functions ranging in $[1 .. m]$ and $[1 .. l]$ respectively.
The \textit{false positive probability} for this scheme is given by:
\begin{equation} \label{Bh-CBF}
f_{fpr} = \left(1-\Sigma^{h}_{j=0}Pr(X = j)\left(1-\frac{1}{l}\right)^{j} \right)^{k}
\end{equation} 

where $Pr(X = j)$ : 

\begin{equation} \label{Bh_Pr}
Pr(X = j) = \left(^{nk}_{j}\right) \left(\frac{1}{m}\right)^{j} \left(1-\frac{1}{m}\right)^{nk - j}
\end{equation}

\noindent The second scheme presented by Rot et al. is the \textit{VI-CBF scheme}. It is nearly similar to \textit{$B_{h} scheme$} with a difference in the number of counters used, this scheme uses a single variable-increment counter per entry that works exactly like the second counter in the previous scheme. Here, there is a decrease in memory usage since only one variable-increment counter is used for each entry, however, this becomes a drawback because it limits accessing the $B_{h} sequence$ directly which is possible in $B_{h} scheme$ thanks to its first counter (which stores the number elements hashed to a specific entry in the filter). \\
The \textit{false positive probability} of \textit{VI-CBF scheme} is given by: 

\begin{equation} \label{VI-increment}
f_{fpr} = \left(1-\left(1-\frac{1}{m} \right)^{nk} - \frac{L-1}{L}\binom{nk}{1} \frac{1}{m}\left(1-\frac{1}{m} \right)^{nk-1} - \frac{(L-1)(L+1)}{6L^{2}} \binom{nk}{2} \left(\frac{1}{m}\right)^{2} \left(1-\frac{1}{m} \right)^{nk-2} \right)^{k} 
\end{equation}

\noindent Many comparisons were held in \cite{6195563}, one of these comparisons compared the \textit{false positive rates} of \textit{CBF}, \textit{$B_{h}scheme$}, \textit{Improved $B_{h}scheme$} and \textit{VI-CBF scheme} by varying the bits per element value. The results in general shows an improvement in the \textit{fpp} whenever the bits size increases reaching an improvement by factor of 7, while the \textit{VI-CBF scheme} has the lowest \textit{fpp} among all the variants included in the comparison.

\subsection{The Fingerprint Counting Bloom filter (FP-CBF)} The variant \textit{FP-CBF} \cite{PONTARELLI2016304} seeks reducing the \textit{fpp} by introducing a different way of representing the items in the filter. It uses $c$ bits for counting and $d$ bits for storing the fingerprints, and it uses $(k+1)$ hash functions such that the first $k$ hash functions are preserved for updating the counters, and the last one $h_{fp}$ is used to update the fingerprint fields. Therefore, the number of bits used in \textit{FP-CBF} is \((c+d) \times m\) where $m$ is the number of cell in the filter. The insertion of a new item to \textit{FP-CBF} occurs by incrementing the \(h_{i}(e)\) \((1 \leq i \leq k)\) positions in the filter by $1$, while the corresponding fingerprint field is updated by \textit{xoring} $h_{fp}(e)$ with the values stored in the fingerprint field on each cell. \\
Concerning the deletion, an item in the filter is deleted by decrementing all the positions \(h_{1}(e), h_{2}(e), ... ,h_{k}(e)\) by $1$ and performing \(h_{fp}(e) \oplus h_{fp}(e)\) which sets the fingerprint field to $0$. Moreover, to query an item, all the corresponding counters should be checked, and if there is at least one counter of $0$ value, then the item \(e \notin S\) ($S$ is the elements set). If the previous condition is not the case, then the fingerprint field of each counter of value $1$ should be equal to \(h_{fp}(e)\), if so, then the item may exist with a small false positive possibility, otherwise, it is not. \\
The idea behind using the the fingerprints is to minimize the possibility of a false positive such that when the counters of an item have the value of $1$, its corresponding fingerprint field should equal to $h_{fp}(e)$, and when another item $e'$ is mapped to the same position, there won't be a false positive unless \(h_{fp}(e) = h_{fp}(e')\) which occurs with the probability of \(\frac{1}{2^{f}}\). \\
\noindent In a \textit{FP-CBF} filter of $m$ cells and $k+1$ hash functions and $n$ stored elements, the false positive probability is given by:
\begin{equation}
    \left(1-\left( 1-\frac{1}{m'} \right)^{kn} - \left( \frac{2^{f}-1}{2^{f}}\right)\frac{1}{m'}\binom{kn}{1}\left( 1-\frac{1}{m'} \right)^{kn-1} \right)^{k}
\end{equation}

\noindent where \(m^{'} = \left( \frac{c}{c+f}\right) \)

\begin{figure}[!h]
  \begin{center}
  \includegraphics[width=0.65\textwidth]{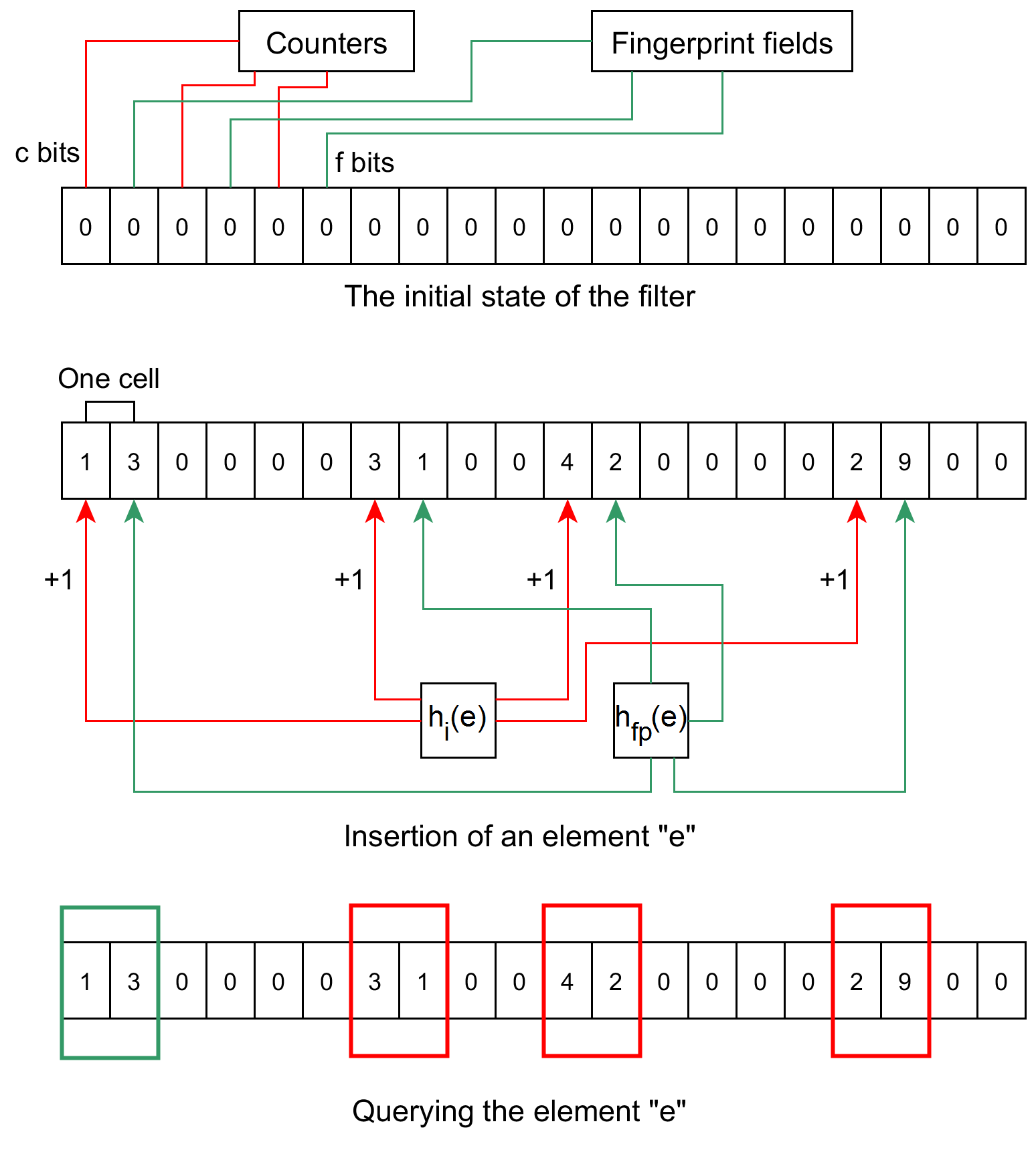}
  \end{center}
  \caption{An example of inserting an element $e$, the cells $0, 3, 5$ and $8$ are incremented by $1$ while the corresponding fingerprint fields on each field are updated by \textit{xoring} \(h_{fp}(e)\) with the target fingerprint fields. For querying the element $e$, only the fingerprint of cell $0$ (inside the green box) is checked since its counter value is $1$, if the result of \textit{xoring} it with \(h_{fp}(e)\) is $0$ then $e$ exists, otherwise $e \notin S$}
  \label{fig:FP-BF}
 \end{figure} 

\subsection{Retouched Bloom Filter}
In the Standard Bloom Filter, the structure is vulnerable to present false positives but no false negatives, besides that, although increasing the bit array size and obtaining a better false positive rate, still have another inconvenience which is the increase in memory space usage. The \textit{Retouched Bloom Filter} (RBF) structure \cite{Donnet:2006:RBF:1368436.1368454} suggests decreasing the false positives by allowing an acceptable amount of false negatives. This is achieved by applying a process called \textit{Bit clearing} which resets the bits corresponding to the false positive, therefore, the new false positive rate becomes: 
 \begin{equation}
 f^{'}_{P}=f_{P}.d_{1}
 \end{equation} such that 
 \begin{equation}
 d_{1}=(1-\frac{1}{p_{1}m})^{k}
 \end{equation} 
 where \(p_{1}\) is the probability of a bit to be set to $1$, $k$ is the number of hash functions and $m$ is the bit array size. Moreover, the probability that an item from the data-set becomes a false negative after the bit clearing process is given as \(\Delta=f_{N}=1-d_{1}\). \\
 Four-bit clearing process variants have proposed to achieve an effective false-positive bits deletion in which one method guaranteed a ratio fraction of the reduced false positives proportion according to the proportion of generated false negatives \(\Sigma = \Delta \frac{f_{P}}{f_{N}}=1\).

\subsection{Accurate Counting Bloom Filter}
The standard counting Bloom Filter uses an array of counters instead of bits and supports dynamic modifications through insertions and deletions (which standard bloom filter cannot handle). Nevertheless, there still a margin of false-positive probability that can be improved. The Accurate Counting Bloom Filter (ACBF) \cite{ACBF} works on reducing the \textit{fpp} by changing the structure of the counters array and dividing it into multi-level arrays where the first level answers the membership queries and the rest of the vectors are used to calculate the counters of the hashed elements. The core of \textit{fpp} reduction resides in the separation of the \(1^{st}\) level array from the rest levels' arrays and increasing its size. The figure below makes the picture clearer. The structure of the ACBF is organized by using the \textit{offset indexing}, therefore, to reach a specific counter, the proposed algorithm of ACBF gets the index of that counter by using a \textit{popcount($b_{j}$, index)} where $b_{j}$ is the array in level $j$ and \textit{index} is the position of the hashed item in the previous level, after that, what is left is to aggregate the bits encountered in each level i.e. if the counter in a level $j$ equals to $1$ then add it to the item's counter, else, the algorithm just stops the counters' spanning.\\
 The optimal false positive probability is given as follows: \((1-(1-\frac{1}{4m-kn})^{nk})^{k}\approx (1-e^{\frac{-kn}{4m-kn}})^{k}\) where \(4m-kn\) is the optimal maximized first level size, $m$ is the number of counters, $n$ is the number of items and $k$ is the number of hash functions.
  \begin{figure}[!h]
  \begin{center}
  \includegraphics[width=0.5\textwidth]{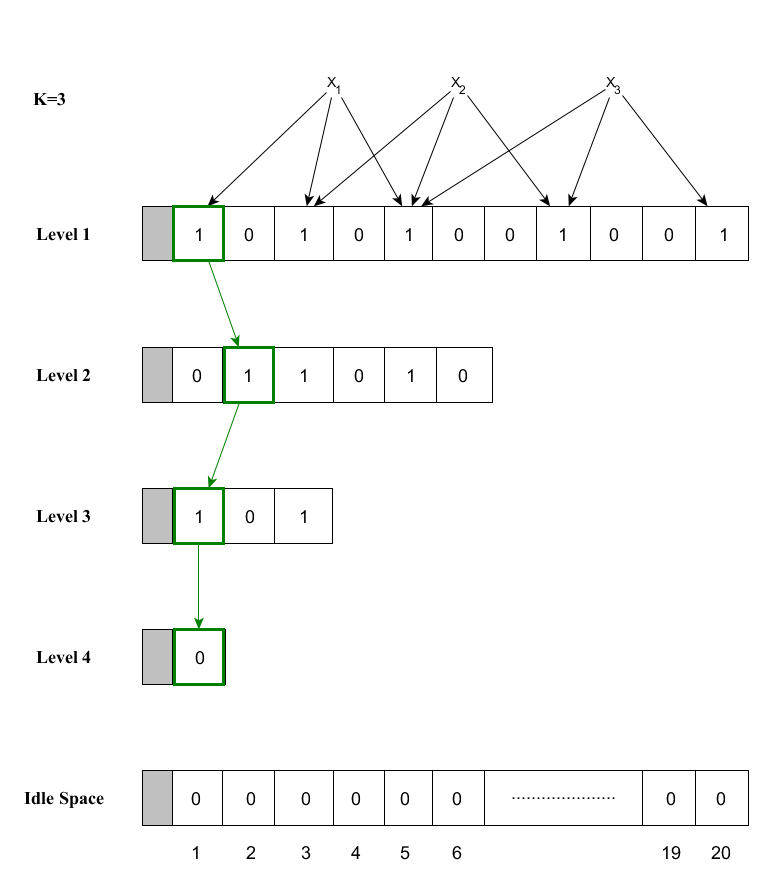}
  \end{center}
  \caption{Accurate Counting Bloom Filter's structure with 4 levels}
  \label{fig:AccurateCBF}
 \end{figure}

\subsection{Generalized Bloom Filter}
The GBF \cite{Laufer2007GeneralizedBF} is another compact data structure that is considered as a modified version of the standard Bloom Filter that allows false negatives to set an upper bound for the false positives. The GBF uses two hash function sets, \(g_{j}(x)\) and \(h_{j}(x)\) where such that ($1 \leq i \leq k_{1}$ and $1 \leq j \leq k_{2}$) such that each function group is responsible for resetting and setting the bits of the filter's array; \(g(x)\) makes the bit equals to 0 and \(h(x)\) makes it 1. Initially, the bits array shouldn't be set to zero, it can be initialized at any value. For the insertion process, all the positions \(g_{k1}(x)\) must be reset i.e 0 and the positions of \(h_{k2}(x)\) must be set to 1. In case of any collision between the two functions \(g\) \& \(h\), the corresponding position will be reset. \\
 The false negatives (x \(\in\) the dataset S but it is considered as not an element of S by the GBF) occur when at least one of the bits corresponding to \(g\) is set (is 1) or one of the bits of \(h\) is reset due to another insertion of an element thereafter. The case of false positives is similar to the standard bloom filter, nevertheless, in GBF, the positions of both \(g\) and \(h\) are 0 and 1 resp. but the item $x$ isn't in the set $S$ due to another insertion of an item. \\

\begin{figure}[!h]
  \begin{center}
  \includegraphics[width=0.5\textwidth]{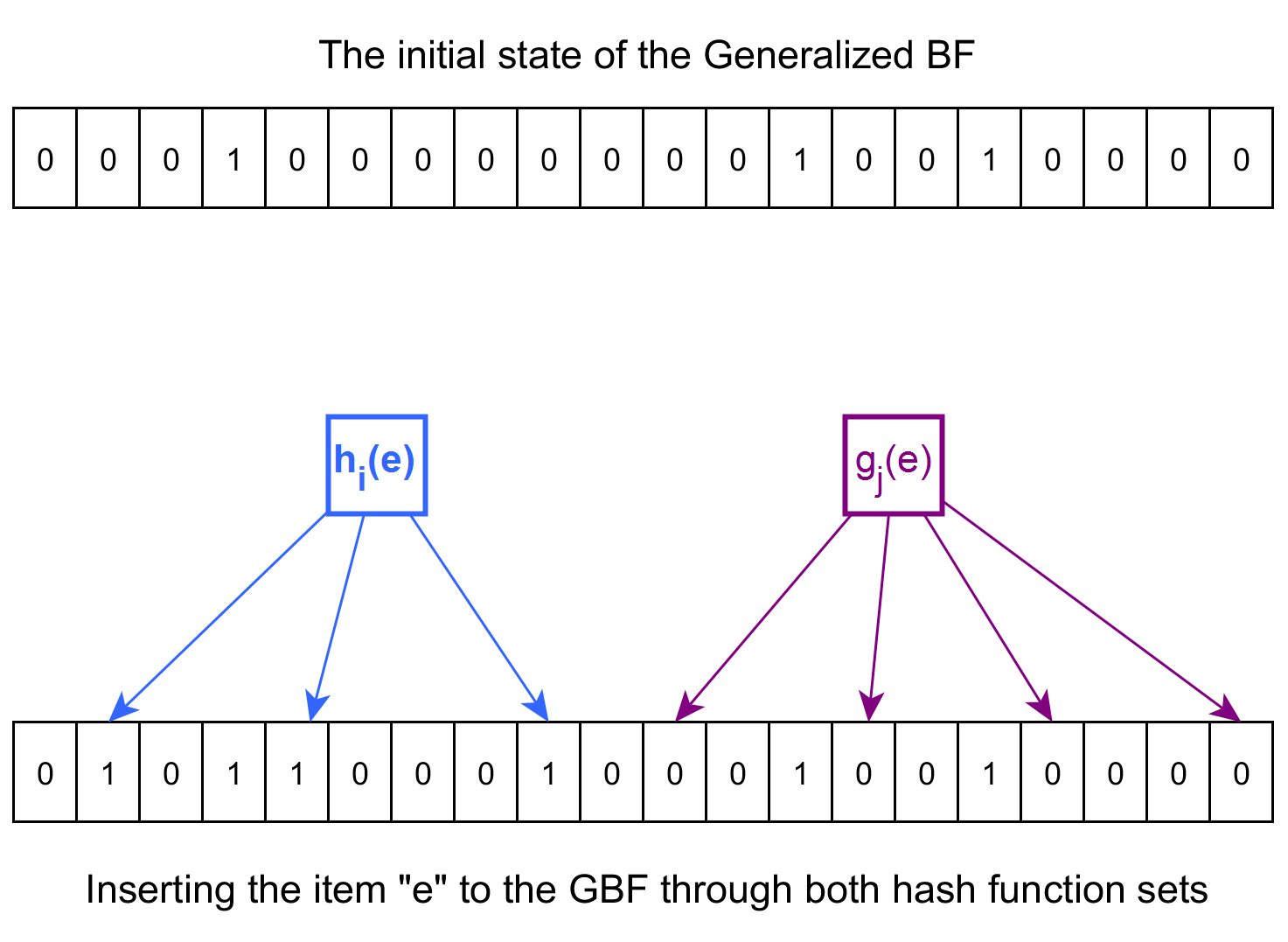}
  \end{center}
  \caption{Insertion of an element to the GBF}
  \label{fig:GeneralizedBF}
 \end{figure}

\noindent The Fig. \ref{fig:GeneralizedBF} is a simple demonstration of the insertion process of an element to the Generalized Bloom Filter of size $m$, the bit vector is not necessarily initialized at $0$, so we can notice that the bits $3, 12$ and $15$ are having the value $1$. Concerning the insertion operation, the hash positions $1, 4$ and $8$ are set to $1$ by the $h_{j}$ hash functions, where the $g_{i}$ functions reset the bits at positions $10, 13, 16$ and $19$.

\noindent To calculate the false positive probability of the Generalized Bloom Filter, the probability that a bit is reset $q_{1}$, the probability that a bit is set $q_{2}$ and the probability that a bit is neither reset not set should be calculated. The probability that a bit is reset means that at least one of the $k_{1}$ hash function of the GBF has reset the bit and it is given by:
\begin{equation} \label{q0}
q_{1} = \left[1 - \left(1 - \frac{1}{m}\right)^{k_{1}}\right] \approx (1 - e^{-k_{0}/m}) 
\end{equation}

Next, the probability that a bit is set means that there are no $g_{i}$ hash function that reset the bit and at least one of the $k_{2}$ hash functions has set the bit, the probability $q_{1}$ is given by:

\begin{equation} \label{q1}
q_{2} = \left[1 - \left(1 - \frac{1}{m}\right)^{k_{2}}\right] \left(1 - \frac{1}{m}\right)^{k_{1}} \approx (1 - e^{-k_{2}/m}) e^{-k_{1}/m} 
\end{equation}

Furthermore, there is another part of the bit vector where the bits are neither set nor reset for each element insertion which hashes the probability of:
\begin{equation}
\begin{aligned}
(1- q_{1} - q_{2})    &= 1 - \left[1 - \left(1 - \frac{1}{m}\right)^{k_{1}}\right] \\
      &  - \left[1 - \left(1 - \frac{1}{m}\right)^{k_{2}}\right] \left(1 - \frac{1}{m}\right)^{k_{1}} \\
      &= \left( 1 - \frac{1}{m}\right)^{k_{1}+k_{2}} \approx e^{-(k_{1}+k_{2})/m}
\end{aligned}
\end{equation}
\noindent Since there is a bit vector of size $m$, then there is an average of $l_{1} = m \times q_{1}$ of bits reset, $l_{2} = m \times q_{2}$ bits set and $m(1 - q_{1} - q_{2}) = (m - l_{1} - l_{2})$ bits which are neither set nor reset.
Finally, the false positive probability can be written as follows:
\begin{equation}
f_{fp} = p^{l_{1}}(1 - p)^{l_{2}}
\end{equation}

\noindent where $p$ is the probability that a specific bit in the vector is $0$ after $n$ element insertions.

\subsection{Multi-Class Bloom Filter}
Since the forwarding table memory space on the low-end switches used in the modern data centers which use multicast is limited and incapable of forwarding a considerable number of group communications concurrently, \cite{6089061} propose using Bloom Filters for each switch interface to encode the multicast groups and therefore use the memory efficiently.
The role of the Bloom Filters is that when a multicast packet arrives, The switch checks which interfaces should forward the packet, however, using BF means that there is a possibility of false positives which causes, in multicast forwarding, what is called the \textit{traffic leakage} which is forwarding the multicast packet from the undesired interface(s).\\
\noindent Before exploring the method to overcome this challenge, let's go over the suggested structure of the Multi-Class Bloom Filter (MC-BF), in a set $S$ there are $N$ elements and the MC-BF size is $m$, moreover, each element $e$ has a probability of being present in the filter called $p_{e}$ \textit{(presence probability)} and a false probability $f_{e}$, from these two metrics, the expected number of elements that were labeled as false positives is given as \(\EX(f_{N}) = (1-p_{e})\times f_{e}\). This structure suggests that the higher is the presence probability of an element $e$ the lower is the expected number of the false-positive elements. \\
\noindent Unlike the standard Bloom Filter, MC-BF uses a set $H_{e}$ of hash functions for each element $e$ in the set $S$, furthermore, Li et al. suggested a method for calculating the number of hash functions for each multicast group such that groups with higher presence probability would have fewer hash functions, on the contrary, the groups with lower presence probability are assigned more hash functions for the sake of reducing the traffic leakage. \\
\noindent The insertion of an element to MC-BF occurs by using the $H_{e}$ hash functions belonging to the element $e$ (in the scenario of multicast forwarding, the hash functions belong to the multicast group) to set the corresponding $k$ bits to $1$ while querying an element requires checking the bits (the same as the standard Bloom Filter) by using the $H_{e}$ hash function set. The false-positive probability can be given as: 

\begin{equation}
f_{e} = [1-(1-\frac{1}{m})^{\Sigma_{i=0}^{N-1}p_{i} \times k_{i}}]^{k_{e}}
\end{equation}

\noindent where $m$ is the filter size, $k_{i}$ represents the number of hash functions of each element $0 \leq i \leq N$, $k_{e}$ is the number of hash functions of the element $e$ and $p_{i}$ is the presence probability of the element $i$.

\begin{figure}[h]
  \begin{center}
  \includegraphics[width=0.65\textwidth]{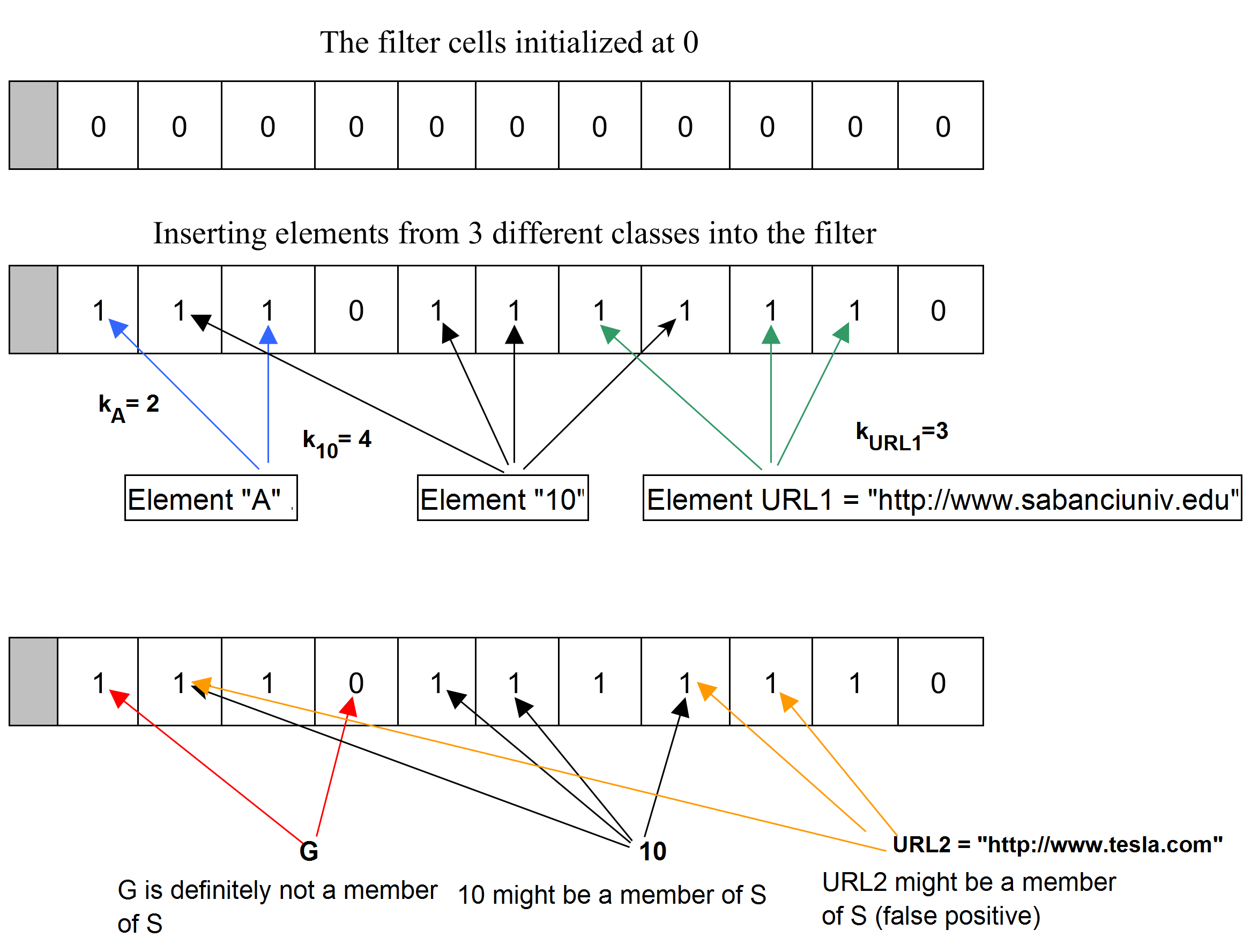}
  \end{center}
  \caption{A simplified MC-BF example with filter size $m = 11$ and $3$ classes (\textit{letters, integers and urls}) where each class of elements has $2, 4$ and $3$ hash functions respectively ($k_{letters}=2$, $k_{integers}=4$ and $k_{urls}=3$).}
  \label{fig:MC-BF}
\end{figure}

\noindent Fig. \ref{fig:MC-BF} demonstrates the operations of insertion and querying, while inserting an element of a class $i$, the $k_{i}$ bits are set to $1$, so when inserting the letter \textit{G}, the bits $0$ and $2$ would be $1$, inserting the number $10$ would set the bits $1, 4, 5$ and $7$ to $1$ and finally adding the URL \textit{"http://www.sabanciuniv.edu"} to the MC-BF makes the hash positions $6, 8$ and $9$ have the value $1$.
\noindent Querying elements existence in MC-BF is quite similar to the query method of the standard Bloom Filter with taking into consideration that each element belongs to the $i^{th}$ class and therefore, only the corresponding hash positions of the hash function set $H_{i}$ should be checked.

\newpage

\subsection{False-positive-free multistage BF}
As one of the networking applications, the Information-Centric Networking - ICN  needs efficient architectures for \textit{Multicast communication} which imposes challenges as accurate forwarding and scaling issues, The \textit{in-packet Bloom Filter} approach minimizes the false positive probability (the router links which are falsely identified for the multicast) which is a crucial problem in the multicast communication (e.g when large database data is forwarded in undesired links in the multicast tree, there will be network overhead that may cause congestion, the same thing goes when a high-quality video is transmitted to links that were falsely and positively matched). The in-packet Bloom Filter \cite{ROTHENBERG20111364} packet has queries for identifying the links that belong to the filter (i.e. the links where the packet should be forwarded) using the bitwise AND and the compare (CMP) operation on the bits that are set of both the link address and the filter. The previous approach is based on a flat structure that stores the multicast tree as a set of edges in a fixed-size filter, however, this method neither annuls false positives nor scales well with the input size. Therefore, János et al. propose the  
\textit{false-positive-free multistage Bloom Filter-MSBF} \cite{6877748} which is composed of stage-Bloom filters of varying sizes. Assuming there is a multicast tree of depth $h$, the MSBF uses $h$ Bloom filters such that the $i^{th}$ Bloom filter of size $m$ is represented by the Elias gamma universal code ~\cite{1055349} such that the packet header is divided into two parts; the first \(\log{m} - 1 \) $\gamma$ bits are for the Elias gamma code and the \(\log{m}\) $\beta$ bits represents the Bloom filter. At each stage, the Bloom filter contains only the information about links at the $i_{th}$ hop, moreover, by benefiting from the previous knowledge of the number of the elements to be added and the ones to be excluded, the filter size may vary to eliminate the chances of having false positives.

\subsection{Complement Bloom filter}
Given a set of elements in the universe $U$ and the set $S$ of elements to be represented by a Bloom filter such that $S = \{x_{1}, x_{2},...., x_{n}\}$ where $n$ is the number of elements in $S$. The Complement Bloom Filter (CompBF) \cite{7264999} proposes the use of an additional set called the \textit{Complement set $S^{c}$} that holds the complement elements of the ones in $S$ to reduce the false-positive answers.
$F(S)$ is the set of positive answers returned by the Bloom filter where for $k$ hash functions, all $h_{i}(e)=1$ where $1 \leq i \leq k$ and $F(S)-S$ stands for the false-positive answers. Therefore, it is possible to write 
\begin{center}
$F(S) = S \cup F(S)-S$.
\end{center}

\begin{figure}[h]
  \begin{center}
  \includegraphics[width=0.65\textwidth]{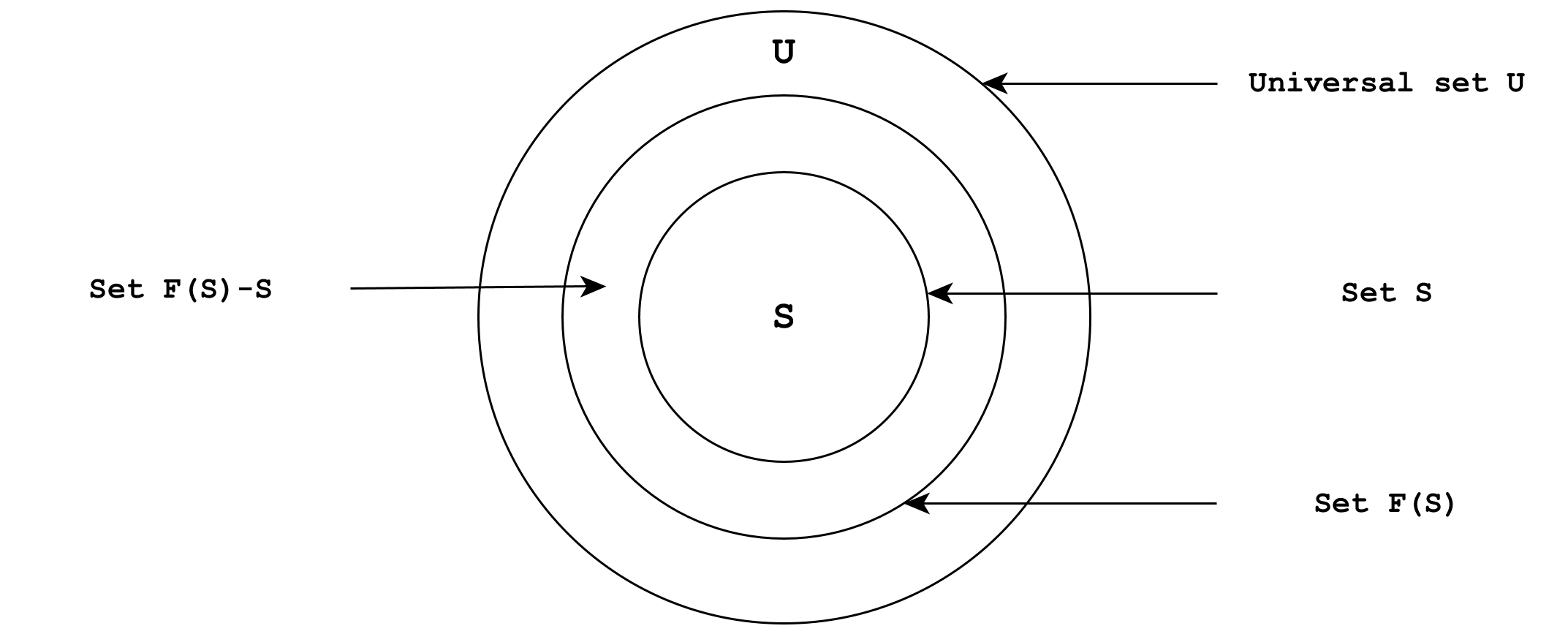}
  \end{center}
  \caption{The sets representation for U, S and F(S) in a standard BF}
  \label{fig:CompBF}
\end{figure}

\noindent Consequently, Lim et al. preferred using another Bloom filter to keep track of the elements of $S^{c}$, and this new set is in use for the Bloom filter positive answers validity checking purpose such that when the additional Bloom filter of the set $S^{c}$ is queried for an element $e$, the negative answer referring to the non-existence of the $e$ means a true positive in the set $S$. \\
\noindent Lim et al. divides the universal set $U$ into two sets $S$ and $S^{c}$, also, it can be split into $F(S)$ and $F(S)^{c}$ where $F(S)$ is the disjunction of $F(S)-S$ and $S$, in the other hand, $F(S^{c})$ is the disjunction of $F(S^{c})-S^{c}$ and $S^{c}$. In both sets $S$ and $S^{c}$, $F(S)-S$ and $F(S^{c})-S^{c}$ stand for the false-positive answers for each set respectively. Therefore, two equations are concluded:

\begin{equation} \label{compBF1}
    F(S) \cap F(S^{c}) = (F(S)-S) \cup (F(S^{c})-S^{c})
\end{equation}
and secondly:
\begin{equation} \label{compBF2}
    (F(S)-S) \cap (F(S^{c})-S^{c}) = \emptyset
\end{equation}

\noindent Eq. \ref{compBF1} means that a false positive answer would come from either one of the sets and not both of them, as Eq. \ref{compBF2} is showing, and in this case, the use of an off-chip hash table is needed to decide the element $e$'s membership.\\  
\noindent Now, for the false positive probability of the whole filter (CompBF), there are several notations and equations to go over in order to arrive to final \textit{fpp} of CompBF. We have $T$ elements in the universal set $U$ where $S$ has $n$ and $S^{c}$ has $n_{c}$ elements respectively, thus, the probability of set $S$ denoted by $P(S)$ is equal to $\frac{n}{T}$ and $\frac{n_{c}}{T}$ for the $P(S^{c})$ where $T = n+n_{c}$.
Let's give the false positive probability for the second Bloom filter of an element $e \notin S^{c}$ but $e \in S$ and let's denote it as $f_{c}$:
\begin{equation} \label{compBF3}
    f_{c} = \left( 1 - \left( 1 - \frac{1}{m^{c}}\right)^{k_{c}n_{c}}\right)^{k_{c}}
\end{equation}
\noindent $f_{c}$ is also considered as the conditional probability given a set $S$, $P(F(S) \mid S)$. (The same counts for the first Bloom filter of set $S$ and it is given by $f$).
Then, we can write the following:
\begin{equation} \label{compBF4}
    P(F(S)-S) = P(S^{c}) \times f
\end{equation}

\begin{equation} \label{compBF5}
    P(F(S^{c})-S^{c}) = P(S) \times f_{c}
\end{equation}
\noindent Then, from Eq. \ref{compBF1} and Eq. \ref{compBF2}:

\begin{equation} \label{compBF6}
    P(F(S) \cap F(S^{c})) = P(F(S)-S) + P(F(S^{c})-S^{c})
\end{equation}

\noindent And since $P(F(S)-S)$ and $P(F(S^{c})-S^{c})$ represent the false positive probabilities of both Bloom filters of the sets $S$ and $S^{c}$ respectively, then:

\begin{equation} \label{compBF7}
    P(F(S) \cap F(S^{c})) = P(S^{c}) \times f + P(S) \times f_{c} 
\end{equation}
Finally, as Eq. \ref{compBF7} shows, this is the conclusion that the false positive probability of \textit{CompBF} is the sum of both filters' false positive probabilities.

\section{Space-efficient Bloom Filters}

\subsection{d-left Counting Bloom Filter (dl-CBF)}
As counting Bloom filter achieves a better memory usage (space-efficient) by allowing more false positives, dlCBF \cite{Bonomi:2006:ICC:1276191.1276252} aims at obtaining more space-saving by applying another hashing function called \textit{d-left}. \\
\textit{dlCBF} is a counting Bloom filter using a d-left hashing and a structure that looks like a bit array. This hashing method is about dividing the bit array into \textit{d}  subtables where each subtable contains \textit{b} buckets where the total number of buckets is \textit{B} and the number of buckets for each subtable is \(\frac{\textit{B}}{\textit{d}}\). Each cell in a bucket contains a fingerprint and a counter, and the fingerprint consists of a \textit{bucket id} and a \textit{remainder} For inserting new incoming items, some uniform choices of buckets from each subtable must be done and then the element is inserted in the corresponding $d$ buckets with the least number of items. In the case of a tie, we choose the bucket with the leftmost subtable. So far, the odds that a false positive occurs happen only when $H(x) = H(y)$ where $x \in S$ and $y \notin S$, and in addition to that, this method is flawed as that when an element $y$ is inserted after the insertion of another element $x$, there is a possibility that both elements have the same remainder, therefore, when a user or an application intends to delete $x$ from the \textit{dlCBF}, it would be confusing whether the remainder belongs to $x$ or $y$, especially that the remainder would be stored in $x's$ bucket in two different positions, so, deleting both copies would lead to a false negative.\\
\noindent A solution for the latter issue, according to Bonomi et al. is using a hash function of two phases, the first phase is to obtain the fingerprint $f_{x}$ (\(f: x \leftarrow [B] \times [R]\)) where $R$ is the range of the remainders, and in the second phase, additional permutations \(P_{i}f_{x}\) for \(1 \leq i \leq d\) are used in such a way that when an element is inserted, a check over the existence of the bucket $b_{i}$ and remainder $r_{i}$ is performed, if it exists already, then the corresponding counter is incremented and the chances for a collision occurrence are less probable.
 
 \begin{figure}[!h]
  \begin{center}
  \includegraphics[width=0.7\textwidth]{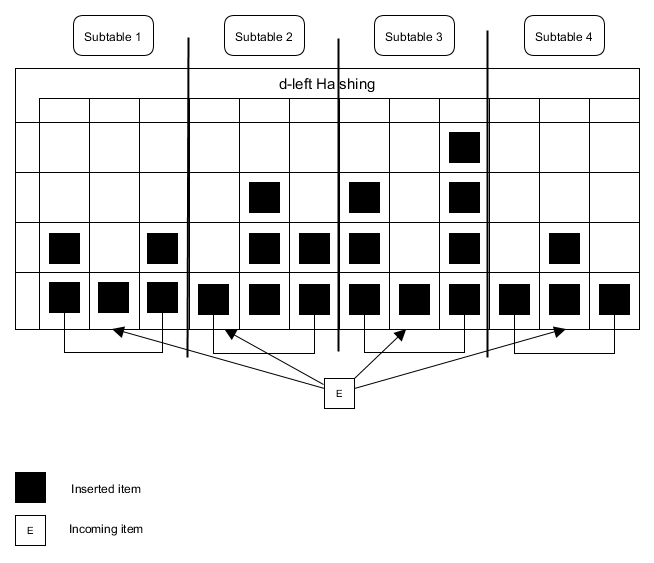}
  \end{center}
  \caption{d-left Hashing}
  \label{fig:d-left BF}
 \end{figure}

\subsection{Memory-optimized BF}
The Memory-Optimized Bloom filter or as Ahmadi et al. call it, the Bloom filter with an additional hash function (BFAH) \cite{4698251} addresses the excessive memory usage in the conventional Bloom filter where it stores $c$ copies of an item and only one copy is used after the lookup operation.
BFAH, suggests the usage of an additional hash function that selects one memory address where the item is to be stored.\\

\noindent When an item $e$ comes, it will be hashed to $k$ slots using $k$ hash functions in the bit-array where each cell points to a memory address. After the hashing, the process to select one physical memory address to store $e$ is performed by the additional hash function $k'$. This operation maps the hashed item to one memory address which avoids creating $c-1$ unnecessary copies by the standard Bloom filter.
\noindent It is important to note that this approach does not annul completely the false positives.

\begin{figure}[!h]
  \begin{center}
  \includegraphics[width=0.9\textwidth]{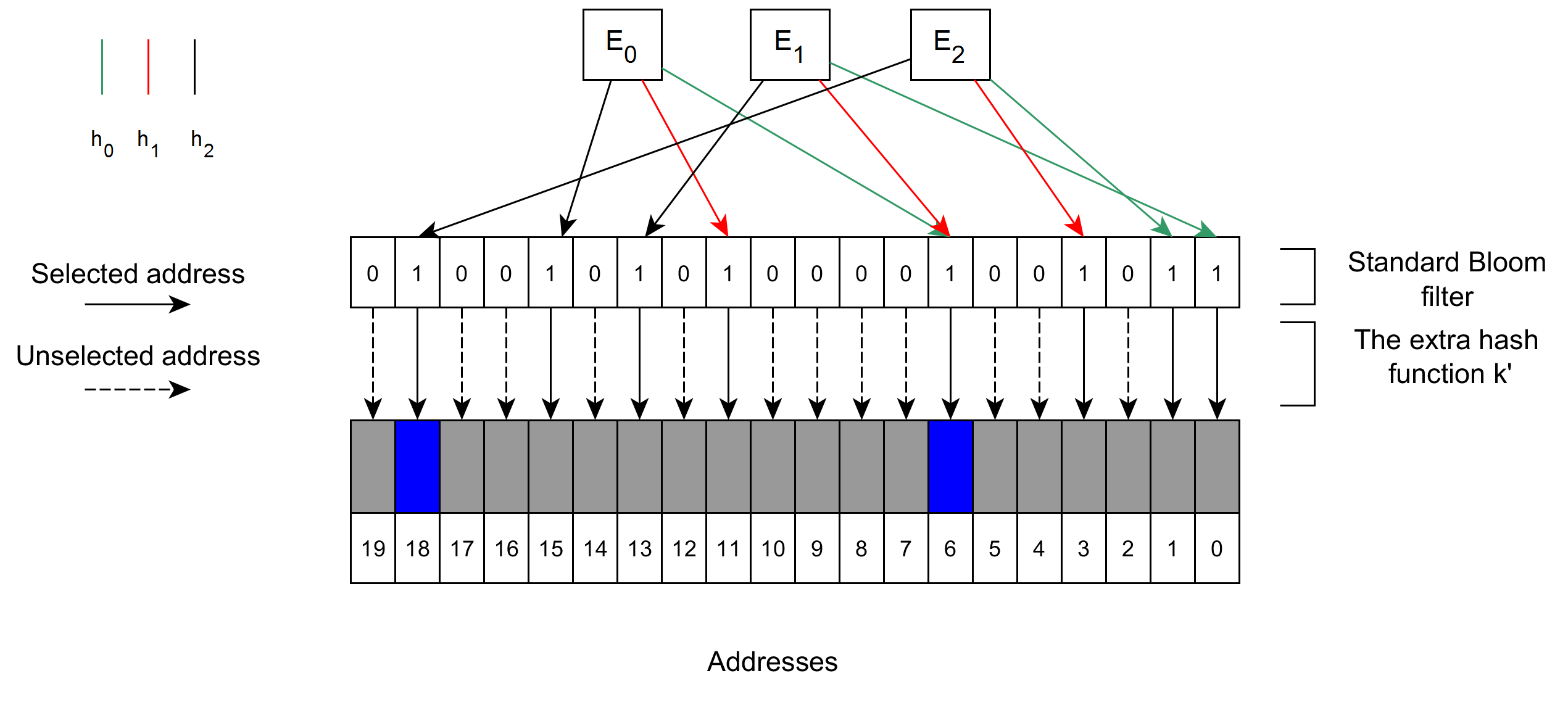}
  \end{center}
  \caption{An example showing the process of inserting $3$ items to the filter where a standard Bloom filter is used in the middle of the process}
  \label{fig:BFAH1}
 \end{figure}

\noindent Fig.\ref{fig:BFAH1} shows the insertion process to the BFAH filter. To insert the element $E_{0}$, $3$ hash functions are used \(h_{0}, h_{1} \) and $h_{2}$ to insert the element to the middle Bloom filter, and the extra hash function $k'$ is responsible for mapping the element to the selected memory address. For example, when inserting $E_{0}$, the array locations $6$, $11$ and $15$ are set to $1$ then by using $k'$ which is a simple hash function (\textit{element's index mod k}). Therefore, the hash value refers to the memory address $6$ where the element to be stored (Slot in Blue in Fig.\ref{fig:BFAH1}). The same for element $E_{1}$ and the array locations $0$, $6$ and $13$ are set to $1$, however, after using the extra hash function, a collision can be detected (false positive) where the selected memory address is the same as the one selected by $E_{0}$ which is $6$.

\begin{figure}[!h]
  \begin{center}
  \includegraphics[width=0.9\textwidth]{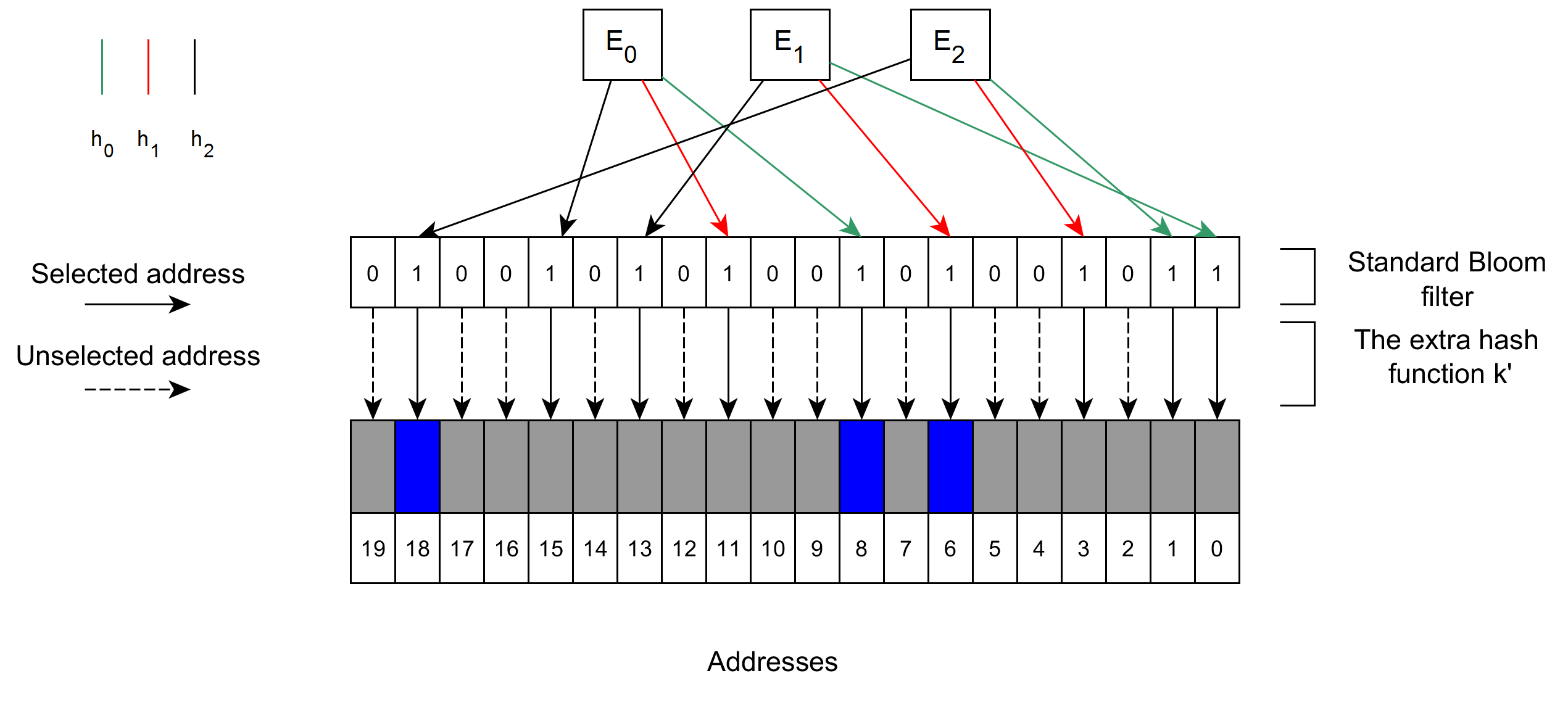}
  \end{center}
  \caption{Another example of insertion without false positives}
  \label{fig:BFAH2}
 \end{figure}

\noindent Fig. \ref{fig:BFAH2} shows a case where there is no false positive, the address location $8$ is selected for element $E_{0}$ and another address $6$ for the element $E_{1}$.

\subsection{Matrix BF}
The Matrix BF (MBF) \cite{6036819} aims to detect copy-paste similarities between documents in a database. The Matrix Bloom filter is an array of rows where each row is a Standard Bloom filter and represents a single document. The MBF has two main operations, insertion and similarity detection. For insertion, It retrieves a document $D_{i}$ from a database and divides it into chunks of sub-strings according to $4$ chunking styles so the similarity detection system would be chunking style-independent. After the splitting process, the MBF hashes all the chunks using $k$ hash functions and sets the corresponding bit locations in the document $D_{i}$'s Bloom filter to $1$. This process is run for all the documents desired to be in the MBF.

\begin{figure}[!h]
  \begin{center}
  \includegraphics[width=0.9\textwidth]{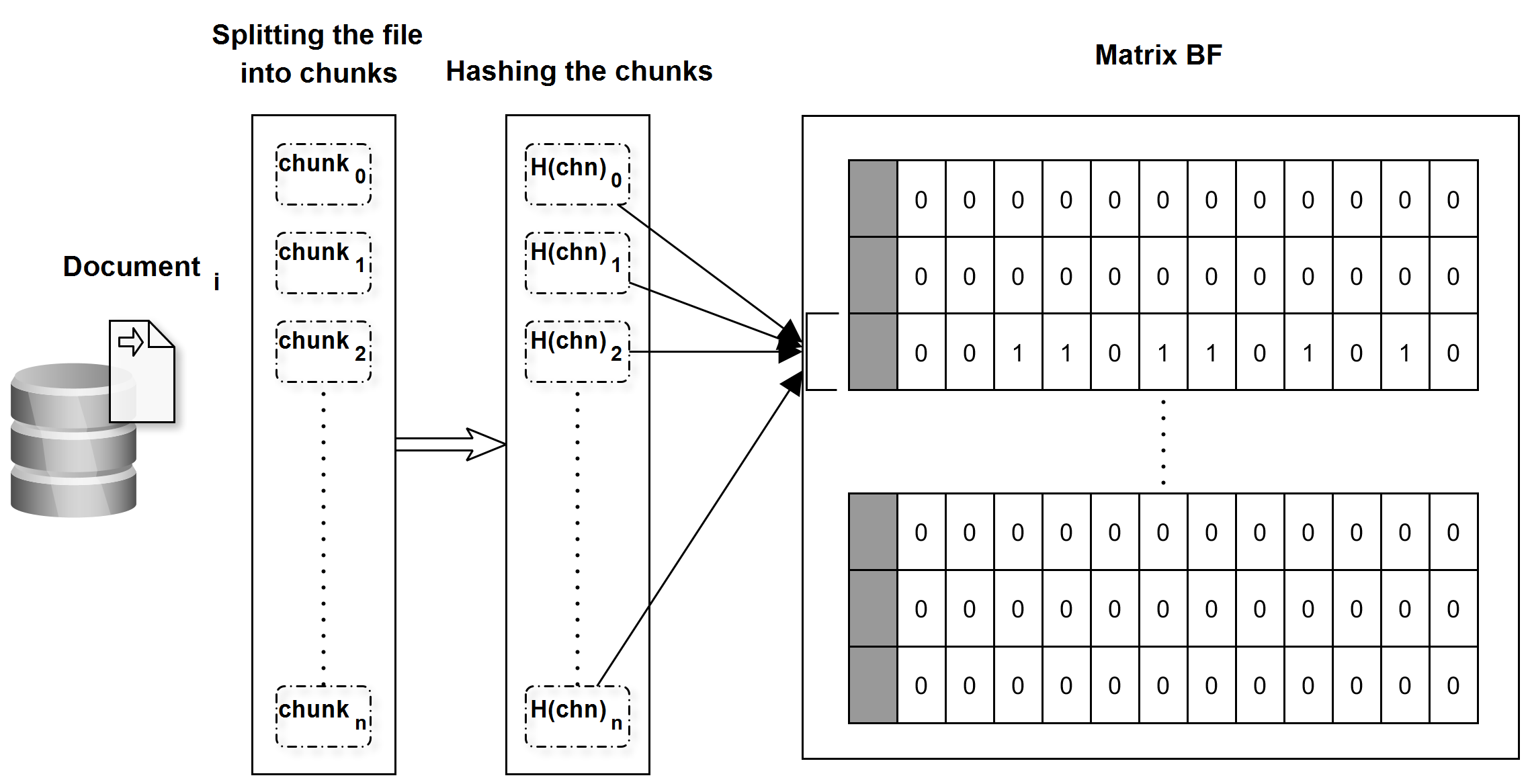}
  \end{center}
  \caption{Splitting the $i^{th}$ document into chunks then hash them to their bit positions in the $i^{th}$ Bloom filter of the matrix}
  \label{fig:MBF-insert}
 \end{figure}

\noindent Detecting copy-paste similarity is almost similar to the insertion phase, for instance, when a document $D_{i}$ is to be queried, then it will be split into chunks and then hashed using the predefined hash functions, and these two operations are the same in the insertion. However, to compare this document with another or many documents, the MBF performs the $AND$ operation bit by bit on each pair of document's bit-array, and the result would show the similarity degree between the documents. At the end of this operation, if the number of resulting $1$s is greater than a predefined similarity threshold then it is highly probable similar to the document(s) that is in comparison with.

\begin{figure}[!h]
  \begin{center}
  \includegraphics[width=1.0\textwidth]{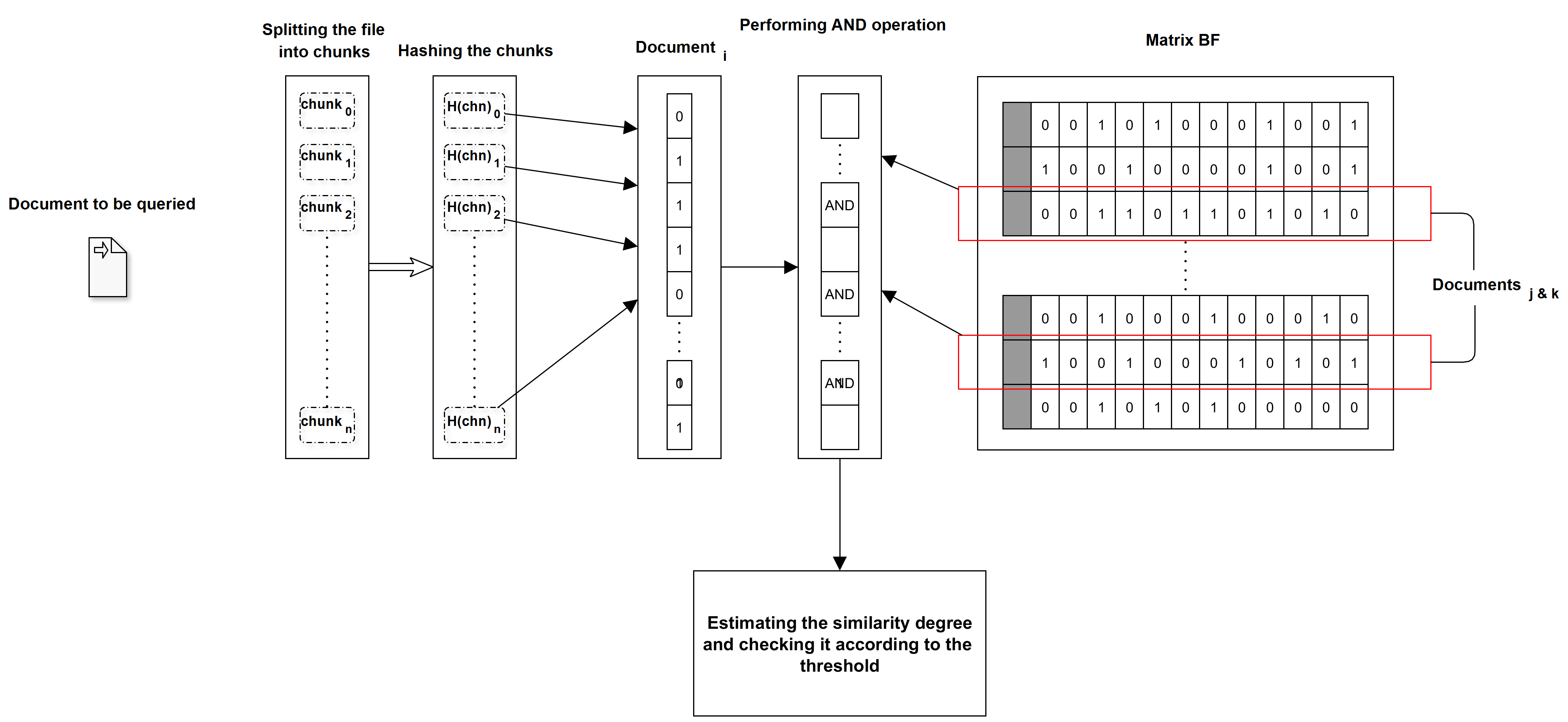}
  \end{center}
  \caption{The process of detecting similarity between document $i$ and documents $j$ and $k$}
  \label{fig:MBF-insert}
 \end{figure}

\subsection{Forest-Structured Bloom Filter}

Since the standard Bloom filter is placed in the RAM, it is put in a memory restriction due to RAM limited space. The Forest-Structured Bloom filter (FBF), as an improved version of the flash-memory Bloom filters, is introduced to get rid of the RAM space restriction by depending on the flash memory-based Solid State Drive (SSD). \\
\noindent FBF divides the SSD space into sub-FBFs and organizes them into $k$ layers wherein each layer there are $\delta$ sub-FBFs grouped into physical blocks in the flash memory, the FBF structure is organized in Forest shape. Each block from a layer $i$ has $c$ children except for the leaf ones. \\
\noindent In case the dataset size fits the RAM, there will be no need to move to the flash memory, and only the first layer with the highest number of blocks is located in the RAM with similar functionality as the conventional Bloom filter. On the other hand, if the dataset size exceeds the RAM capacity then the FBF is moved to the SSD and $c$ children blocks are added to each block so they form a new lower layer, there is a capacity, $C$ for each layer, once it is reached, a new layer is formed by inserting more block children of the previous layers' blocks. Consequently, the RAM will be used as a buffer during the insertion process. \\
\noindent For querying an element $e$ in FBF, it is based on several hash function groups, first, by applying $h_{0}$ on the element $e$, the \textit{block\_id} is identified. Second, $h_{1}(e)$ helps to identify which sub-BF is responsible for the element $e$. And last, a set of $k'$ independent hash functions are used to check the bit positions in the sub-BF, if all are $1$ then $e$ exist in the filter, otherwise, one of the children blocks of the current block (where $e$ was checked and was not found) is selected to pursue the lookup process in it, the next block is selected by applying: \\

\begin{equation} \label{eq:FBF-nextLoc}
\begin{aligned}
    block\_id &= block\_id \times c + num\_block\_root \\
    block\_id &= block\_id - (h_{1}(e) \gg length - layer_{parent} \times \log_{2} c) \% c
\end{aligned}
\end{equation}

\noindent Following equation \ref{eq:FBF-nextLoc}, the next sub-BF's index is located and therefore the lookup process continues from there. If the process keeps giving negative responses about the existence of the element $e$ then it is deduced that $e$ is not a member of the filter and needs to be inserted for further potential queries.

\subsection{Compacted Bloom Filter}
As the network's bandwidth must be saved to the maximum during transmissions between the network nodes, in some cases, the Bloom filters may impact the network especially when it is needed to send them back and forth between the nodes frequently. The Compacted Bloom filter (CmBF) \cite{7809719} uses less space and therefore is faster while transmitting over the network.
\noindent To generate a CmBF, the original Bloom filter is divided into $k$ blocks, each of $n$ elements. The CmBF is an array of $n$ indices where each index is of $m$ bits. Each $m$ bits on each index represents the bit pattern in the standard Bloom filter, in other words, \textit{index\_$i$} in the CmBF stands for the $i^{th}$ bit from each block in the Bloom filter. 
\noindent The method of assigning index values for each index in the CmBF is by checking the bit patterns in the standard Bloom filter and see whether the pattern $S_{i}$ contains $1$ or not, therefore, $4$ cases can be distinguished:
\begin{itemize}
    \item if $S_{i}$ contains no $1$'s then $CmBF[i] = 0$ (in binary)
    \item if $S_{i}$ contains only one $1$ then $CmBF[i] = r$ where $r$ is the index of the block in which the corresponding bit was set to $1$.
    \item if all the bits of $S_{i}$ are $1$s or half or more than the half of the bits of $S_{i}$ are $1$s then $CmBF[i] = 2^m - 1$ 
    \item if less than the half of bits of $S_{i}$ are $1$s then a random block which contains the $i^{th}$ bit position which is $1$ and assign its index as the index value of CmBF i.e. $CmBF[i] = t$ where $t$ represents the index of the block in the standard Bloom filter.
\end{itemize}
This method introduces more false positives and false negatives too since in the $3^{rd}$ the CmBF considers the bits which are set to $0$ as $1$s (increasing false positives rate), and in $4^{th}$ the filter considers the bits which are set to $1$ as $0$s, as a result, it increases the false negatives rate.
\noindent One solution proposed by the authors is that they introduced a new rule to make use of all the index values in the mapping from bit patterns to the CmBF indices. The new rule states that if the bit pattern value $Si_{val} < 2^m - 1$ then the index in CmBF is assigned the bit pattern's value.
Concerning the query in CmBF, after it is sent over the network, a new standard Bloom filter can be constructed based on the CmBF and the membership query is performed on the bit patterns in the newly formed Bloom filter.
\noindent The CmBF has occupied less space than the standard Bloom filter and has a lower false-positive rate, however, it still introduces false negatives.

 \begin{figure}[!h]
  \begin{center}
  \includegraphics[width=0.8\textwidth]{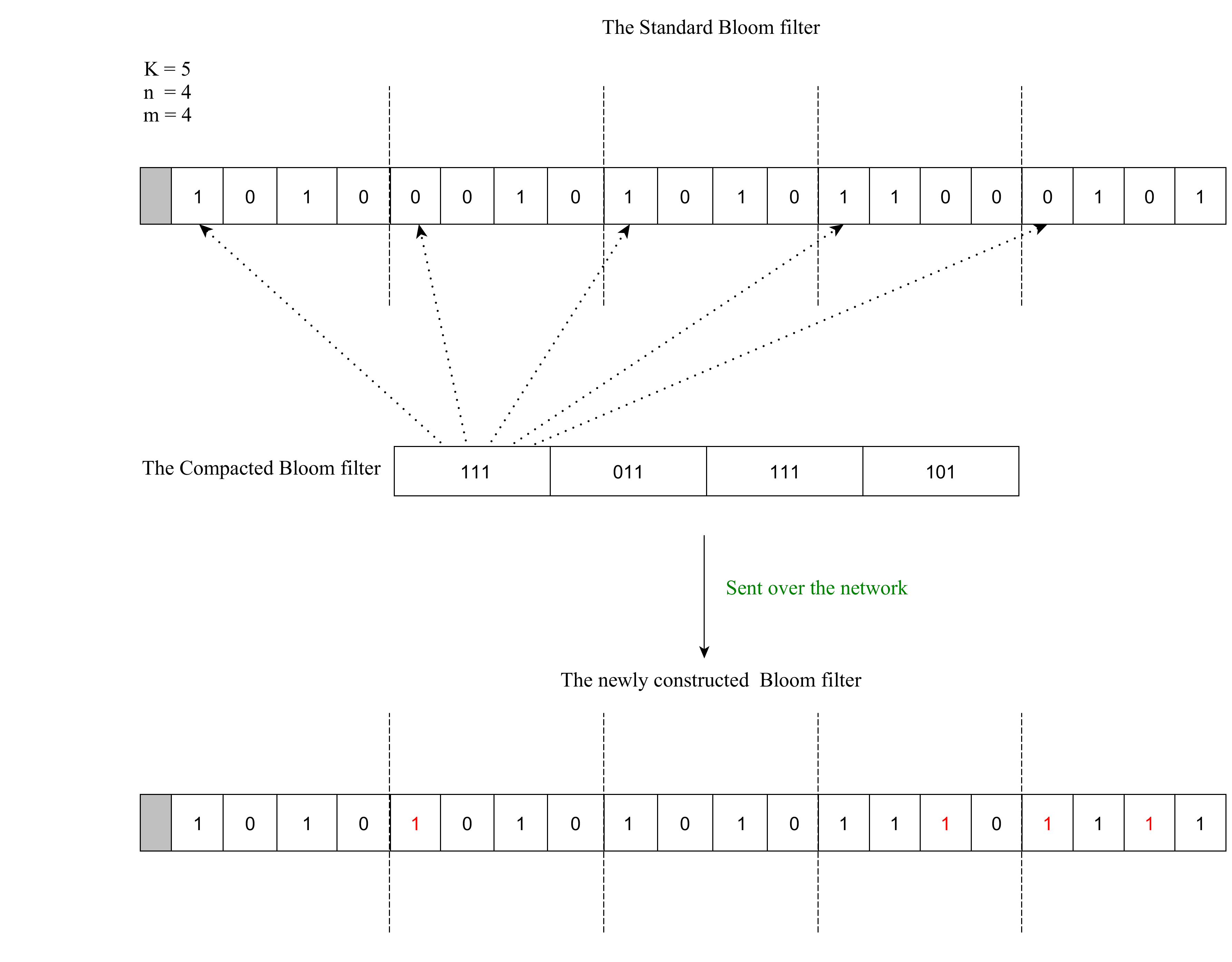}
  \end{center}
  \caption{An example showing the process of converting the standard Bloom filter to a Compacted one and send it over the network, then rebuilding it again. It appears that in the newly constructed Bloom filter, it altered 4 bits from $0$ to $1$ (the bits marked in red)}
  \label{fig:CmBF}
 \end{figure}

\section{Computationally-optimized Bloom Filters}

\subsection{One-Hashing Bloom Filter}
\textit{One Hashing Bloom Filter} (OHBF) \cite{OHBF} is another variant of the standard Bloom filter, the key idea about OBHF is that it uses only one hash function plus some other operations to reduce the computations overhead caused by using $k$ hash functions. 
In a nutshell, the OHBF divides the bits vector of the standard bloom filter into chunks or partitions $m_{i}$ where $k=i$, each partition is of different size, during the insertion process, the item is hashed using only one hash function $h(e_{j})$ then the algorithm applies a modulo operation \(h(e_{j})\mod m_{i}\) so that it will flip the corresponding bit inside the $m_{i}$ partition into $1$.

 \begin{figure}[!h]
  \begin{center}
  \includegraphics[width=0.7\textwidth]{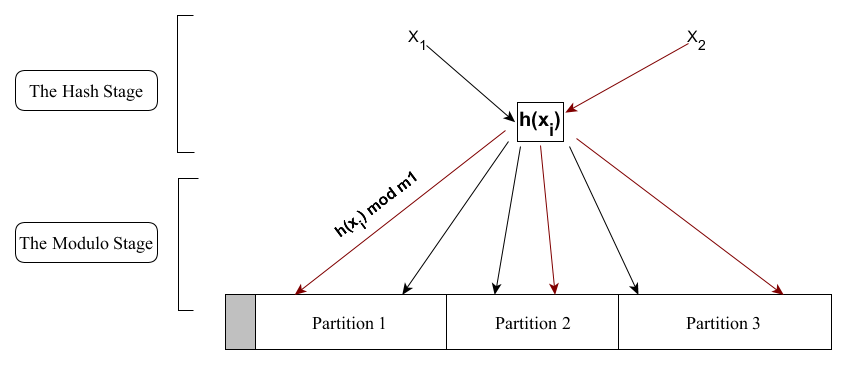}
  \end{center}
  \caption{One-Hashing Bloom Filter showing two stages of hashing and mapping to different partitions of the bits vector where k=3, n=2, and only one hash function}
  \label{fig:OHBF}
 \end{figure}

\subsection{Ultra-Fast Bloom Filter}
The Ultra-Fast Bloom filter (UFBF) is a compact randomized data structure that supports a quick membership query that copes with high-speed network links. The filter consists of $l$ blocks where each block is a set of $k$ words of length $w$, thus, the whole size of the UFBF equals to: 
$m = l \times k \times w$. \\
\noindent To insert an element in the Ultra-Fast Bloom filter, first, a block is randomly selected using the hash function $h_{0}$ ($h_{0}(x) = block_{i}$), after that, the rest of the $k$ hash functions are used to set the corresponding bit locations to $1$. These bit-locations are selected from each word in a way that each word is associated with a hash function ($h_{i}(x) = word[i]$) where $1 \leq i \leq k$. \\
\noindent Querying an element in UFBF is quite similar to the insertion where first, the targeted block is selected using $h_{0}$ then the picked bit-locations are checked whether they are set or reset. If all are set then the query is answered positively, otherwise, it does not exist in the set $S$.

\section{Bloom Filters dealing with Multisets}

\subsection{Spectral Bloom Filter}

The work of \cite{Cohen:2003:SBF:872757.872787} introduces a variant of the Bloom filter, the Spectral Bloom filter (SBF). It pays the trade-off of using more memory space than the standard Bloom filter, furthermore, it handles insertion/deletion operations over the datasets. SBF presents a small change in the original structure where it uses counters instead of bit flags which is nearly similar to the Counting Bloom filter. 
The Spectral Bloom Filter is considered a similar structure to the Counting Bloom Filter (CBF) where both can increase and decrease the $h$ counters of the elements, therefore, the filter is considered as fully supportive for both insertions and deletions. The slight difference between \textit{CBF} and \textit{SBF} is that the latter uses only the minimum $i^{th}$ counter as an estimator for an item's frequency. Furthermore, the \textit{SBF} optimizes the used data structure by increasing only the minimum $i^{th}$ counter among all the $h$ counters, which improves further the accuracy of an item frequency estimation. 

 \begin{figure}[!h]
  \begin{center}
  \includegraphics[width=0.9\textwidth]{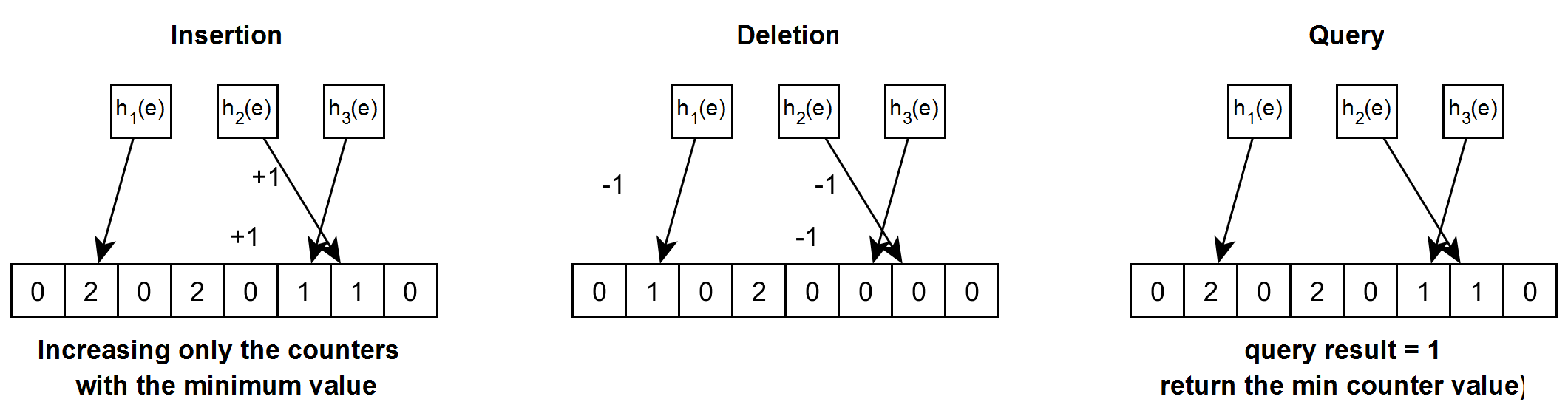}
  \end{center}
  \caption{Insertion/Deletion/Query operations for Spectral Bloom filter}
  \label{fig:spectralBF}
 \end{figure}

\subsection{Adaptive Bloom Filter}
The Adaptive Bloom filter (ADF) \cite{AdaptiveBF} can be considered as a variant of the Counting Bloom filter, in addition to the counting feature that allows extracting the inserted items' frequencies by assigning a counter to each item, it implements an incremental dynamic hash functions creation. The key idea is to insert the incoming item into the filter as the standard Bloom filter also does, then, the ABF calculates a new hash value using the \(k^{th}+1\) hash function (assuming that initially there are $k$ hash functions) and set the corresponding location in the bit array to $1$ and so on, it continues at the same pace until it reaches the \(k+N+1^{th}\) hash function wherever the bit is 0. \\
During the query process, the filter responds to the membership query and returns $N$ as the number of occurrences of the item under the condition that the used hash functions are independent and the collisions rate is zero \cite{AdaptiveBF}. \\ 
The improved ABF is an enhanced version of ABF where $2$ more bits are added to the filter to reduce the error rate of reporting the items' frequencies.
The simulations were run on a \textit{zipf's} distribution and the trials showed that the ABF can be used in unpredictable datasets.

\subsection{The Shifting Bloom Filter}
The Shifting Bloom filter \textit{(ShBF)} \cite{10.14778/2876473.2876476} is a probabilistic data structure that processes the Membership, Association and Multiplicity set queries and aims to use less memory than what the \textit{Standard Bloom filter} uses. Moreover, the main goal for this structure is to store auxiliary information about members set within the filter without needing additional memory, as it were to be in the \textit{Standard Bloom filter}. \\
The basic idea behind \textit{ShBF} is record the item existence information and an additional one concerning the item, for example, the item's multiplicity or type. So, when the item is inserted using $k$ independent hash functions in a bit array of size $m$, the $k$ bits \(h_{0}(x)\mod m, h_{1}(x)\mod m, .... h_{k}(x)\mod m\) are set to $1$, furthermore, the bits at \((h_{0}(x)+o(x))\mod m, (h_{1}(x)+o(x))\mod m, .... (h_{k}(x)+o(x))\mod m\) are also set to $1$ which represent the auxiliary information of the inserted item. \textit{o(x)} stands for the offset of the element. However, in order to decrease the number of memory accesses, \(\frac{k}{2}\) bits will be used to store the existence information, and the other \(\frac{k}{2}\) bits to store the auxiliary information such that for each element \(x \in S\), the bits at \(h_{1}(x)\mod m,.... h_{\frac{k}{2}}(x)\mod m\) and the other bits at \(h_{1}(x)\mod m+o(x),.... h_{\frac{k}{2}}(x)\mod m+o(x)\) are set to $1$. While querying an item, if all the $k$ bits are set to $1$ then \(x \in S\) otherwise \(x \notin S\). 
\\
The \textit{false positive rate} for \textit{ShBF} is given by:
\begin{equation}
f_{fpr} \approx (1-p)^{\frac{k}{2}} \left(1-p+\frac{1}{\bar{w}-1}p^{2}\right)^{\frac{k}{2}}
\end{equation}
where \(\bar{w}\) is a function of machine word size, and \(p = e^{\frac{-nk}{m}}\) such that $n$ is the number of elements in the filter, $m$ is the filter's size.

\section{Bloom Filters dealing with dynamic sets}

\subsection{Dynamic Bloom Filter Sketch}
 The Standard Bloom filter treats the membership queries of datasets that are considered static and of a fixed size whereas, in real-life examples, these datasets are updated frequently either by inserting or deleting items. The fact that the Bloom filter cannot deal with these dynamic datasets makes it an imperfect structure. Therefore, a new data structure called \textit{Dynamic Bloom Filter (DBF)} was suggested by \cite{Guo:2010:DBF:1685872.1685987} to deal with both static and dynamic datasets and to handle insertion and deletion operations. 
 \textit{DBF} is actually a set of \textit{b} \textit{Standard Bloom Filters}. During the insertion process, an incoming item is inserted inside the \textit{ActiveSBF}, which is an SBF that still has enough capacity to handle new items. In the case that inserting a new item to the SBF is not possible, the \textit{Dynamic Bloom Filter} will create a new active SBF and hash the item into it, then increments the number of the active SBFs $b$ by $1$. The DBF performs the deletion process by: 
\begin{enumerate}
\item Identifying the SBF containing all the \(h_{i}(x)\) hashes of the item $x$, where they must be all set to a non-zero value. If there is only one value set to $0$, then the deletion will be aborted. If the hashed values are included in many SBFs, then the process will be stopped since it will be impossible to be certain that the erased item is the same one to be deleted.
\item After deleting $x$, a \textit{Merge} operation is called to replace two active SBFs with their union in case of their union doesn't exceed one SBF's capacity: \textit{\(activeSBF.n_{1}\)}+\textit{\(activeSBF.n_{2}\)} \(\leq C\), where $C$ is the capacity of one SBF. 
\end{enumerate}
Before going through the \textit{False Positive Probability}, it is important to note that the \textit{Dynamic Bloom Filter} is a set of \textit{Standard Bloom Filters}, and here, the \textit{SBF} is meant to be the \textit{Counting Bloom Filter} since it has the property of deleting items from the filter. Hence, it is comparable to \textit{DBF} (i.e. both have the insertion/deletion features). The \textit{False Positive Probability} can be one of two formulas depending on the cardinality of the set $N$ and an \textit{SBF's} capacity $C$:

\begin{enumerate}

\item The \textit{fpp} of the \textit{DBF} is the same as a \textit{SBF's} probability given previously in equation (\ref{fpp_SBF}) if the cardinality of the set is greater than the capacity of an \textit{SBF} from the \textit{DBF}.
 
\item Otherwise, the false positive probability of the whole \textit{DBF} depends on the probabilities of its \textit{SBFs}; therefore, the \textit{fpp} of the first \(p-1\) \textit{SBFs} is denoted by \(f_{m,k,C'}\) and \(f_{m,k,N_{last}}\) such that \(N_{last} = N - C \times [N/C]\) represents the \textit{fpp} of the last \textit{SBF}. Thus, the probability that not all counters of the \textit{SBFs} are not zero can be given as:

\begin{equation} \label{fpp_DBF}
\begin{split}
f_{m,k,C,N} &= 1 - (1 - f_{m,k,C,C})^{[N/C]} (1 - f_{m,k,C,N_{last}}) \\
			&= 1 - (1 - (1 - e^{-k \times C/m})^{k})^{[N/C]} \\
			&  1 - (1 - e^{-k \times (N-C \times [N/C])/m})^{k})
\end{split}
\end{equation}    

Here, $m$ stands for the number of bits in the bit vector, and $k$ represents k independent hash functions.

\end{enumerate}

\subsection{Weighted Bloom Filter}
The Weighted Bloom Filter \cite{WBF} suggests a method of a dynamic assignment of hash functions to each element of the set that is going to be inserted into the filter. The way of determining the number of hash functions to be used for each item is decided by its query frequency \(f_{e}\) and its probability of being a member \(x_{e}\) of the set $S$. The formula of the Weighted Bloom Filter's false positive probability, which is the sum of each element's false probability, is given as:
\begin{equation} 
P_{fp}=\Sigma_{e\in U}r_{e}.(1-p)^{k_{e}}
\end{equation} 
where $e$ is the element, and 
\begin{equation}
r_{e}=\frac{(1-X_{e})f_{e}}{\Sigma_{i \in U}(1-X_{i}) \cdot f_{i}}
\end{equation} 
$X_{e}$ represents an indicator of an element's existence, $1$ shows that the element $e$ exists and $0$ for the opposite. $r_{e}$ is the normalized query frequency of the element $e$. Here $k_{e}$ denotes the $nk$ in the standard Bloom filter. \\
Moreover, it was shown that the Weighted Bloom Filter is a generalization of the standard bloom filter considering the case where both $f_{e}$ and $x_{e}$ are constant which means that the query frequency and the membership likelihood are similar for all the items in the universe $U$.\\
A drawback of the Weighted Bloom Filter is that the value of $k_{e}$ has to be calculated before each query process such that if an element is being queried frequently then $k_{e}$ has to be calculated over and over which causes more computational overhead. 

\subsection{Invertible Bloom Lookup Tables}
This variant of the standard Bloom filter \textit{(IBLT)} \cite{DBLP:journals/corr/abs-1101-2245} supports an additional operation other than what the standard Bloom filter provides which is listing the pairs it contains as long as the number of these pairs $(x,y)$ does not exceed a predefined threshold $\rho$. Concerning the structure of the \textit{IBLT}, it is divided into $m$ cells in which the cells are organized in $k$ sub-tables, each of size $m/k$ to guarantee that the hashes are mapped into a distinct location. The insert process is straightforward so whenever a new pair is inserted, three fields corresponding to a specific cell will be modified as follows:  \\
1) \(T[h_{i}(x)].count+=1\)  \\
2) \(T[h_{i}(x)].keySum+=x\) \\
3) \(T[h_{i}(x)].valueSum+=y\) \\
where this operation is repeated for all $k$ hash functions used in the \textit{IBLT}. The delete operation is simply performed by applying the subtraction instead of addition while there is the \textit{GET} operation which is the lookup process, it is similar to the one in the standard bloom filter. On the other side, the \textit{LIST\_ENTRIES} operation is responsible for extracting all the entries in the cells which have non-zero counts. At the end of this process, if Table $T$ is empty then the output list is the whole set of entries in the \textit{IBLT}, otherwise, this method has only output a partial list of the entries (key-value pairs) in the \textit{IBLT}.

\section{Bloom Filters with multi functions and purposes}
\subsection{Counting Bloom Filter}
In the standard bloom filter, we usually use $d$ hash functions to map each item to its cells in the bit array we have for satisfying membership queries, however, this technique suffers from a drawback, standard bloom filters cannot support deletions, therefore, \cite{Fan00summarycache:} suggested a variant of bloom filter which is CBF (Counting Bloom Filter). CBF uses counters instead of bits so when a new item comes, the counter is incremented by $1$ and decremented by $1$ in case of deletion. 
\noindent For memory concerns, \cite{Fan00summarycache:} suggested $4$ bits per counter and considered as sufficient for many applications. In previous sections, we have shown some variants of the Counting Bloom filter.
 
 \begin{figure}[!h]
  \begin{center}
  \includegraphics[width=0.9\textwidth]{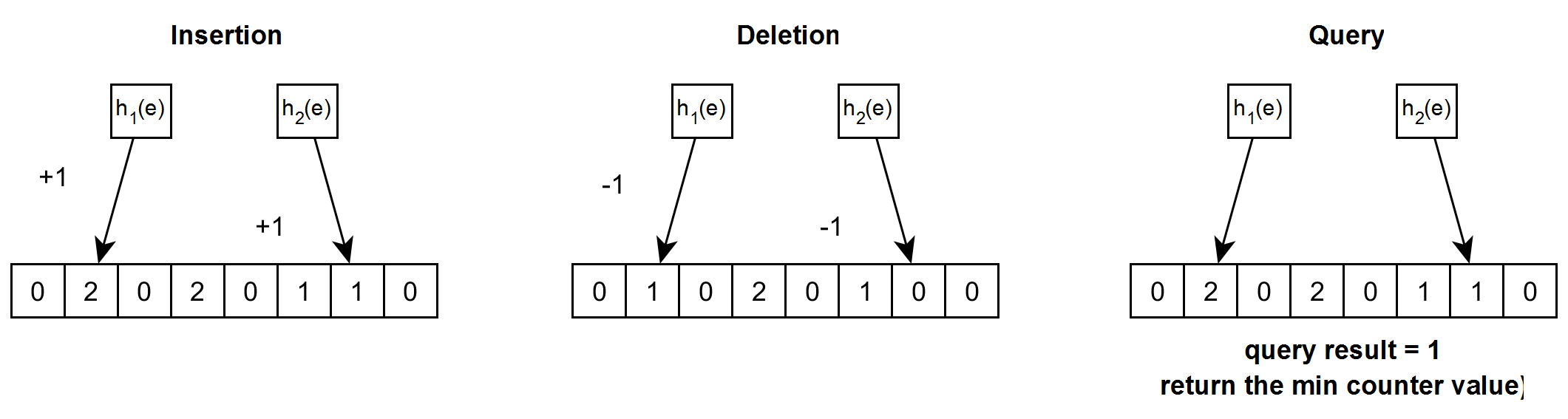}
  \end{center}
  \caption{Insertion/Deletion/Query operations for Counting Bloom filter}
  \label{fig:countingBF}
 \end{figure}
 
\subsection{Deletable Bloom Filter}
The idea behind the Deletable Bloom filter \cite{DlBF} is simply to guarantee the prevention of false negatives occurrences while deleting items from the filter which is achievable only by paying a trade-off with memory space where there is a filter of size $m$ divided into $k$ regions and a part of the filter's space is reserved for encoding the regions where the collisions occurred. Therefore, the insertion process will have a slight change after hashing an element to different bits using $d$ hash functions, after insertion, when a collision occurred, then the region which witnessed the (bits overlapping) will be encoded in the extra space as a non-deletable region indicating which region is the intended one. Furthermore, during the removing process, an item is deleted by resetting its $k$ bits to 0 if they are in collision-free-regions (this information can be extracted from the extra space). \\
 The probability of an item being deletable is given by: 
 \begin{equation}
 p_{d}=(1-(1-p_{c})^{\frac{m^{'}}{d}})^{h}
 \end{equation} 
 where; $h$ is the number of hash functions, $d$ is number of regions obtained after dividing the filter's bits array, \(m^{'}\) is the size of the filter without the region responsible for encoding the collision-free-regions and \(p_{c}\) is the probability that at least a collision occurs in one bit cell given by \(p_{c}=1-p_{1}-p_{0}\). \\
 The previous implementation showed that the increase in the extra used memory space for regions encoding has effect on the deletion process and false positive rates.
 
\begin{figure}[!h]
  \begin{center}
  \includegraphics[width=0.9\textwidth]{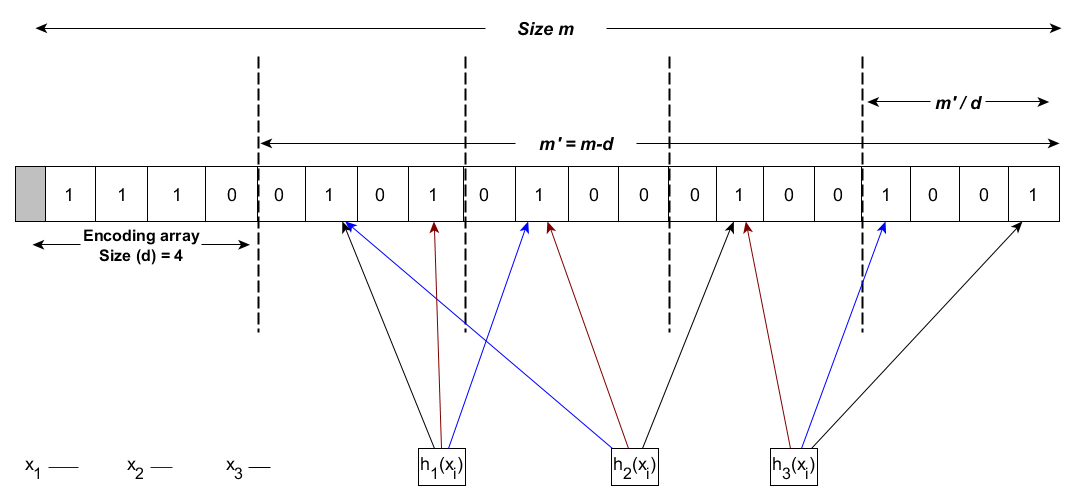}
  \end{center}
  \caption{Deletable Bloom filter's structure}
  \label{fig:deletableBF}
 \end{figure}

\subsection{Distance-Sensitive Bloom Filters}
\begin{figure}[!htbp]
  \begin{center}
  \includegraphics[width=0.2\textwidth]{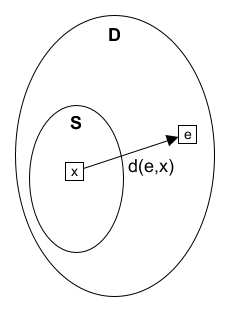}
  \end{center}
  \caption{A figure showing the distance calculated by the metric \(d\)}
  \label{fig:d-sensitive-BF}
 \end{figure}

The Distance-Sensitive Bloom filter (D-SBF) \cite{DBLP:journals/corr/GoswamiP0S16} is a variant of the standard BF which answers queries of the form "Is element \textit{e} close to an element of the sub-set \textit{D} of the original data stream?". The basic element of which this filter is built on is the locality-sensitive hashing function and it aims at differentiating between elements \textit{e} where \(d(e,x) \leq \epsilon\)  and elements \textit{e} where \(d(e,x) > \delta\) and the parameters \(0 \leq \epsilon < \delta\), \((d\) is the distance metric) supposing that we have a finite set \(S\subset D\) and \(x\in S, e\in D\). which means that distance-sensitive bloom filter allows a margin of both false positives and false negatives. The case where (\(\epsilon = 0\)) represents the standard bloom filter with no false negatives. Figure \ref{fig:d-sensitive-BF} shows the distance calculated as $d$.

\subsection{Persistent Bloom Filter}
As mentioned before in the \textit{Standard Bloom Filter}, this compact data structure answers membership queries about the existence of items in a given data sets, however, one of the shortcomings of this structure is that it responds only to whether an item exists or not, and it doesn't cover temporal queries which ask for the existence of some actions or items depending on a time interval or it expensively treats this sort of queries. Persistent Bloom filter \textit{PBF} was designed and proposed by \cite{Peng:2018:PBF:3183713.3183737} to cover temporal queries and it is considered as a data structure for temporal membership testing queries. It is composed of neatly selected SBFs where each standard bloom filter is responsible for a subset of items distributed according to time intervals. PBF decomposes a single temporal membership testing query (tmt-query) into $n$ standard membership testing queries and distributes them to the SBFs. 
Peng et al. addressed the problem of the \textit{fpp} getting increased to the \textit{n}-th power since a tmt-query of length $n$ well be decomposed into $n$ tmt-queries to an \textit{SBF}. Therefore, they proposed a data structure which complies better than the standard one which is \textit{PBF} with its two forms \textit{PBF-1} and \textit{PBF-2}.     
 
\subsection{Cuckoo Filter}
Although the Bloom filter is useful for answering the membership queries for items in a given data streams with minimum false positives, however, it is still not efficient in dealing with deletion operation, \cite{Fan:2014:CFP:2674005.2674994} presents Cuckoo Filter which provides a solution to deal with deletions in Bloom Filters. 
\noindent Cuckoo filter insertion process is based on using two hash functions \(h_{1}\) \& \(h_{2}\) and an array of buckets, when an item $x$ is going to be inserted, if there is at least an empty bucket corresponding to $x$ then a fixed-size fingerprint of the item $x$ \footnote{\label{cuckoo_insert}Hashing the items into fixed-size fingerprints before adding them to the hash table for space efficiency purposes} is added to the free bucket and the process is completed, otherwise, if both buckets were occupied by previously added items, then the algorithm applies a \textit{"relocating process"} where the item occupying that bucket will be relocated to another free bucket and the new item $x$ is placed in the newly freed bucket. \\
 An essential detail concerning the insertion process, when it is needed to get the item $x$ to be relocated while adding another item, and since the items' fingerprints are used, therefore, the relocating process would be hard because the item $x$ cannot be recovered. Therefore, Fan et al. used a method called \textit{partial-key cuckoo hashing} which uses the item $x$'s fingerprint in finding the candidate buckets. The equation (\ref{cuckoo_xor}) makes use of the property of the \(\oplus\) operator such that the new location where the item to be relocated is calculated without getting the original item and hashing it.

\begin{equation} \label{cuckoo_xor}
\begin{split}
h_{1}(x) &= hash(x) \\
h_{2}(x) &= h_{1}(x) \oplus hash(fingerprint(x))
\end{split}
\end{equation}

And the new location is calculated this way:
\begin{equation}
j = i \oplus hash(fingerprint(x))
\end{equation}

Concerning the deletion, it is a basic operation where the Cuckoo filter checks both buckets for the existence of the item to be deleted, if there is a match with the fingerprint of the item in either of the buckets, that fingerprint is removed from the bucket. 

As the Cuckoo filter claims to be a space-efficient structure that outperforms the standard Bloom filter, it is important to highlight the optimal parameters which provide the most efficient space usage, therefore, we can give the space efficiency formula as the following:

\begin{equation} \label{space-eff_eq}
C = \frac{table\_size}{Num. of items} = \frac{f \cdot \{num. of entries\}}{\alpha \cdot \{num. of entries\}} = \frac{f}{\alpha}
\end{equation}

Where:
\begin{itemize}
\item \textit{f}: number of bits for each fingerprint
\item \textit{$\alpha$}: the hash table's load factor
\item \textit{C}: the space cost required for each item
\end{itemize}

Furthermore, some factors play an important role in reducing \textit{C} to the minimum, knowing which value of \textit{fpp} is chosen and the best possible bucket size \textit{b}.
Fan et al. introduced an additional technique to improve the space usage called \textit{Semi-sorting buckets}.  

\subsection{High-Dimensional Bloom Filter}
As it has been seen before in the Standard Bloom filter  \cite{Bloom:1970:STH:362686.362692}, it maps each item from a set into a bit vector with a false positive probability, the High-Dimensional Bloom Filter (HDBF) \cite{info9070159} extends mapping strings into bit arrays to hashing more high dimensional data like vectors into counter arrays using a modified hash function which discretizes vectors of high numerical dimension (in the paper it is the \texttt{sax\_hash function)}. More formally, the HDBF can be represented as: \\
Initially, the HDBF has a counter array where all counters are set to 0, and $k$ hash High Dimensional Integer Hash function (HDIH) (intuitively independent hash functions) $h_{i}$. The dataset is a set of multi-dimensional $n$ vectors \(S={V_{1}, V_{2}, V_{3}, .....,V_{n}}\). When a vector of the set $S$ is mapped in to the HDBF with $h_{i}$ then its corresponding counter positions \((h(V_{i})\mod m)\) will be increased by $1$ ($m$ is the array size). Concerning querying the HDBF, by going through all $k$ HDIH functions, if the \(h(V(q)_{i})\) are greater than 1 then $q$ is an element of the set $S$ with an FPP (false positive probability), otherwise, it is definitely not. The false positivie probability is given as:
 \(f_{HDBF}=(1-p)^{k}=(1-e^{\frac{-kn}{m}})^{k}\) where $p$ is approximately the probability of a one counter remaining $0$.

\section{Conclusion}
The growth of the recent applications in machine learning \cite{rae2019metalearning}, networks, and other several fields is increasing so fast and the variants of the Bloom filter are relatively exploding since the data sources, sizes, and flow are related to recent domain applications such as Deep learning, high-performance computing, privacy-preserving algorithms, and other domains. \\
In this survey, we tried to present variants of Bloom filters and support them with figures explaining the way it answers the membership queries using less or more memory space than what the \textit{Standard Bloom filter} uses, more specifically, showing how each work balances the trade-off between the accuracy and the memory budget. \\
The work is concluded by Bloom filters classification according to the application domain of each filter. We believe that this survey would help new researchers who are getting more familiar with the Bloom filter, and also those who wish to select a specific filter based on their study's requirements. Moreover, we present a compact comparison between the Bloom filters mentioned in this survey based on their main traits (Counting, False negatives, and Deletion operation) in addition to the type of the filter's result, either a True/False answer or an item/pair frequency. This comparison was inspired by the work of \cite{luo2018optimizing}.

 \newpage

\begin{table}[!h]
\centering
\caption{The main traits of each bloom filter variants-(C)Counting-(D)Deletions-(FN)False negatives}
\label{blooms-features}
\resizebox{\textwidth}{!}{%
\begin{tabular}{@{}cccccc@{}}
\toprule
\textbf{The Filter} & \textbf{Main Trait} & \textbf{C} & \textbf{D} & \textbf{FN} & \textbf{Result} \\
\hline \midrule
Standard Bloom Filter & \begin{tabular}[c]{@{}c@{}}Membership query, \(x \in^{?} S\) \end{tabular} & No & No & No & Boolean \\
\hline
Cuckoo BF & Using cuckoo hashing in bloom filter & No & Yes & No & Boolean \\
\hline
Counting BF & \begin{tabular}[c]{@{}c@{}}Membership query + Item's frequency\end{tabular} & Yes & Yes & / & Boolean  / Frequency \\
\hline
Compressed BF & \begin{tabular}[c]{@{}c@{}}Compressing the filter for transmission \end{tabular} & No & No & No & Boolean \\
\hline
Conscious BF & \begin{tabular}[c]{@{}c@{}}Adapting Num. of hashes of $x$ to its popularity\end{tabular} & No & No & No & Boolean \\
\hline
Dynamic BF & Growing dynamically + Allowing Deletions & Yes & Yes & No & Boolean \\
\hline
Persistent BF & Supporting temporal membership queries & No & No & No & Boolean \\
\hline
Spectral BF & Item frequency queries & Yes & Yes & / & Frequency \\
\hline
D-left Counting BF & \begin{tabular}[c]{@{}c@{}}Membership + freq. queries + d-left hashing\end{tabular} & Yes & Yes & / & Boolean / Frequency \\
\hline
The Bloomier Filter & Frequency \& function value & Yes & No & No & Frequency of functions \\
\hline
Distance-Sensitive BF & Querying the distance to an item of a set & No & No & Yes & Boolean \\
\hline
Generalized BF & 2 groups of Set (1) and Reset (0) hash functions & No & No & Yes & Boolean \\
\hline
High-Dimensional BF & Mapping dimensional data to counter arrays & Yes & Yes & / & Boolean / Frequency \\
\hline
Accurate Counting BF & Mapping items to multi-level counter arrays & Yes & Yes & / & Boolean / Frequency \\
\hline
One-Hashing BF & \begin{tabular}[c]{@{}c@{}}Using one hash function and modulo operations \\ for items mapping\end{tabular} & No & No & No & Boolean \\
\hline
Retouched BF & Allowing false negatives to improve the fpp rate & No & No & Yes & Boolean \\
\hline
Deletable BF & Removing items is based on the probability \(p_{d}\) & No & Yes & No & Boolean \\
\hline
Adaptive BF & Incremental dynamic hash function creation & Yes & No & No & Boolean \\
\hline
Weighted BF & Items get more bits according to its popularity & No & No & No & Boolean \\
\hline
IBLT & \begin{tabular}[c]{@{}c@{}}Difference between two sets + Holding the key\\ and count value for each items\end{tabular} & Yes & Yes & / & Boolean / Frequency \\
\hline
VI-CBF & \begin{tabular}[c]{@{}c@{}}Membership query + Item's frequency\end{tabular} & Yes & Yes & / & Boolean / Frequency \\
\hline
Shifting BF & \begin{tabular}[c]{@{}c@{}}Membership query + Association + Multiplicity \end{tabular} & No & No & No & Boolean \\
\hline
Yes-no BF & \begin{tabular}[c]{@{}c@{}}Membership query + false positive elements \end{tabular} & No & No & Yes & Boolean \\ 
\bottomrule
\end{tabular}%
}
\end{table}

In Table 1, some brief information about the main characteristics of each bloom filter variant (including the standard bloom filter). Besides that, an information about the counting, deletion, false negatives occurring possibility is given under the (C, D, FN) columns. The type of the output of each bloom filter is indicated under the last column (Result). Additionally, it is substantial to point that this table is partially inspired from \cite{luo2018optimizing} and additional modifications and Bloom filter variants were brought to it.

\newpage

\makeatletter
\newcommand\footnoteref[1]{\protected@xdef\@thefnmark{\ref{#1}}\@footnotemark}
\makeatother

\newcommand{\tabhead}[1]{\textbf{#1}}
\begin{landscape}
\begin{table}[!ht]
  \caption{A table stating the distribution of bloom filter variants according to their application contexts - \(1^{st} Part\)}
  \label{tab:bf1}
  \centering
  \begin{threeparttable}
  \renewcommand{\arraystretch}{2}
  \begin{minipage}{\textwidth}
    \begin{tabular}{l l l l }
    \hline
      \toprule
      \tabhead{Sketches\tnote{1}} &
      \tabhead{Networking\tnote{1}} & \tabhead{Databases} & \tabhead{Other}\\
      \midrule
      \hline
      Standard Bloom Filter & \begin{tabular}{@{}l}\cite{Bloom:1970:STH:362686.362692} \cite{Geravand:2013:SBF:2560974.2561537} \\ \cite{doi:10.1080/15427951.2004.10129096} \cite{bloomierInNetworks}\footnote{\label{fnote}Papers collaborating to overlay and peer-to-peer networks} \\\cite{1210033} \footnoteref{fnote} \cite{Ledlie:2002:SPS:1133373.1133397} \footnoteref{fnote}\\ \cite{Hsiao:2001:GRS:509506.509515} \footnote{\label{fnote2}For Resource Routing} \cite{Rhea}\footnoteref{fnote2} \\\cite{1019229} \footnote{\label{fnote3}Packet Routing} \cite{916648}\footnoteref{fnote3} \cite{SINGH2018440}(IoT) \end{tabular} & \begin{tabular}{@{}l}\cite{Bloom:1970:STH:362686.362692} \cite{appDB-BF} \\ \cite{appDB-BF1} \cite{CALDERONI20154}\footnote{\label{pp}Privacy-Preservation} \\ \cite{appDB-CS} \footnote{\label{cs} Content Synchronization} \cite{Chen2014RobustSR} \\ \cite{appDB-CS1}
\end{tabular} &\cite{Bloom:1970:STH:362686.362692}(Big Data) \\
	\hline
      Cuckoo BF & \cite{Fan:2014:CFP:2674005.2674994}        &  &  \\
      \hline
      Count-BF &  \cite{Geravand:2013:SBF:2560974.2561537}   &\cite{appDB-BF}   & \\
      \hline
      Compressed BF &  \cite{Mitzenmacher:2001:CBF:383962.384004} \cite{Geravand:2013:SBF:2560974.2561537}   &   & \cite{Mitzenmacher:2001:CBF:383962.384004} \\
      \hline
      Conscious BF &\cite{appNet-ConBF}  &  &  \\
	  \hline      
      Dynamic BF & \cite{Guo:2010:DBF:1685872.1685987} \cite{Geravand:2013:SBF:2560974.2561537}    &  &  \\ 
	  \hline      
      Persistent BF &\cite{Peng:2018:PBF:3183713.3183737}  &\cite{Peng:2018:PBF:3183713.3183737}  &  \\
	  \hline
      Spectral BF & \cite{Cohen:2003:SBF:872757.872787} \cite{Geravand:2013:SBF:2560974.2561537}  & \cite{Cohen:2003:SBF:872757.872787} \cite{Geravand:2013:SBF:2560974.2561537}  & \cite{Cohen:2003:SBF:872757.872787}(Big Data) \\
      \hline
      d-left Counting BF &  \cite{Bonomi:2006:ICC:1276191.1276252}\cite{doi:10.1080/15427951.2004.10129096}  & \cite{Bonomi:2006:ICC:1276191.1276252}  &  \\      
      
      \bottomrule\\
    \end{tabular}
    \end{minipage}
  \end{threeparttable}
\end{table}
\end{landscape}

\newpage

\begin{landscape}
\begin{table}[!h]
  \caption{A table stating the distribution of bloom filter variants according to their application contexts- \(2^{nd} Part\)}
  \label{tab:bf2}
  \centering
  \begin{threeparttable}
  \renewcommand{\arraystretch}{2}
  \begin{minipage}{\textwidth}
  
    \begin{tabular}{l l l l}
    \hline
      \toprule
      \tabhead{Sketches\tnote{1}} &
      \tabhead{Networking\tnote{1}} & 					\tabhead{Databases} & \tabhead{Other}\\
      \midrule
      
      \hline
      The Bloomier Filter & \begin{tabular}{@{}l} \cite{Chazelle2004TheBF} \cite{doi:10.1080/15427951.2004.10129096} \\\cite{bloomierInNetworks}\ref{fnote} \cite{1210033} \ref{fnote} \\\cite{Ledlie:2002:SPS:1133373.1133397} \ref{fnote} \cite{Hsiao:2001:GRS:509506.509515} \footnoteref{fnote2} \\\cite{Rhea}\footnoteref{fnote2} \cite{1019229} \footnoteref{fnote3} \\\cite{916648}\footnoteref{fnote3}\end{tabular}  & \cite{Chazelle2004TheBF} & \cite{Chazelle2004TheBF} \\
      \hline
      Distance-Sensitive BF &\cite{DBLP:journals/corr/GoswamiP0S16} \cite{Snoeren:2002:SIT:611408.611410} & \cite{DBLP:journals/corr/GoswamiP0S16} &\cite{DBLP:journals/corr/GoswamiP0S16} \cite{using-bloom-filters-refine-web-search-results} \\
      \hline
      Generalized BF &\cite{Laufer2007GeneralizedBF} & & \\
      \hline
      High-Dimensional BF & & & \\
	  \hline      
      Accurate Counting BF &\cite{doi:10.1080/15427951.2004.10129096} &\cite{ACBF} &\cite{MapReduce} \footnote{\label{map}MapReduce is a programming model and an associated implementation for processing and generating large data sets} \\
	  \hline      
      One-Hashing BF &\cite{OHBF} & & \\
      \hline      
      Retouched BF &\cite{Donnet:2006:RBF:1368436.1368454} & & \\
      \hline
      Deletable BF &\cite{DlBF} & & \\
	  \hline      
      Adaptive BF &\cite{AdaptiveBF} & & \\
      \hline      
      Weighted BF &\cite{WBF} & &\cite{appWBF} \\
      \hline      
      Invertible Bloom Lookup Tables &\cite{DBLP:journals/corr/abs-1101-2245} \cite{Bonomi:2006:BBF:1159913.1159950} &\cite{DBLP:journals/corr/abs-1101-2245} \cite{Stonebraker:2010:SDC:1831407.1831411} &\cite{DBLP:journals/corr/abs-1101-2245}(Privacy-Preservation) \\
      \bottomrule\\
    \end{tabular}
  \end{minipage}  
  \end{threeparttable}
\end{table}
\end{landscape}

\bibliographystyle{alpha}
\bibliography{refBL}
\end{document}